\documentclass[amsmath,amssymb,twocolumn,aps,superscriptaddress]{revtex4-1}

\usepackage{graphicx}
\usepackage{dcolumn}
\usepackage{bm}
\usepackage{floatrow}
\usepackage{comment}
\usepackage{soul}
\usepackage[caption=false]{subfig}

\usepackage[table,xcdraw]{xcolor}
\usepackage{amsmath}
\usepackage{graphicx}
\usepackage[colorinlistoftodos]{todonotes}
\usepackage[colorlinks=true, allcolors=blue]{hyperref}

\usepackage[ruled]{algorithm2e}

\usepackage{bbm}
\usepackage{amsfonts}
\usepackage{multirow}
\usepackage{amssymb}
\usepackage{natbib}
\usepackage{enumitem}

\usepackage[english]{babel}

\newcommand{\beginsupplement}{%
        \setcounter{table}{0}
        \renewcommand{\thetable}{S\arabic{table}}%
        \setcounter{figure}{0}
        \renewcommand{\thefigure}{S\arabic{figure}}%
        \setcounter{section}{0}
        \renewcommand{\thesection}{\arabic{section}}  
     }

\usepackage[resetlabels]{multibib}
\newcites{Supp}{SuppReferences}

\begin{document}
\captionsetup[subfigure]{labelformat=empty}

\preprint{APS/123-QED}

\title{Influencing dynamics on social networks without knowledge of network microstructure}

\author{Matthew Garrod}
\affiliation{Department of Mathematics, Imperial College London, London SW7 2AZ, UK}%
\author{Nick S. Jones}%
\affiliation{Department of Mathematics, Imperial College London, London SW7 2AZ, UK}
\affiliation{EPSRC Centre for the Mathematics of Precision Healthcare, Department of Mathematics,
	Imperial College London, London SW7 2AZ, United Kingdom.}%

\date{\today}

\begin{abstract}
Social network based information campaigns can be used for promoting beneficial health behaviours and mitigating polarisation (e.g. regarding climate change or vaccines). Network-based intervention strategies typically rely on full knowledge of network structure. It is largely not possible or desirable to obtain population-level social network data due to availability and privacy issues. It is easier to obtain information about individuals' attributes (e.g. age, income), which are jointly informative of an individual's opinions and their social network position. We investigate strategies for influencing the system state in a statistical mechanics based model of opinion formation. Using synthetic and data based examples we illustrate the advantages of implementing coarse-grained influence strategies on Ising models with modular structure in the presence of external fields. Our work provides a scalable methodology for influencing Ising systems on large graphs and the first exploration of the Ising influence problem in the presence of ambient (social) fields. By exploiting the observation that strong ambient fields can simplify control of networked dynamics, our findings open the possibility of efficiently computing and implementing public information campaigns using insights from social network theory without costly or invasive levels of data collection.
\end{abstract}

\pacs{Valid PACS appear here}
\maketitle


\section{Background} \label{Background}

Effective public information campaigns (e.g. concerning vaccinations or climate change) deliver accurate information to the relevant audiences. One way to identify audiences is based on socio-demographic groups (e.g. based on age, income, ethnicity). A more accurate way to build audiences would be to augment this information with population-level social connectivity data. However, population-level social network information is scarcely used in practice due to substantial privacy, reproducibility and generalizability concerns. Given this, we might ask -- what level of information is required to effectively influence opinion dynamics on networks? And, can we reliably influence large scale social systems using only a coarse-grained privacy preserving summary of the social connectivity?

In a public health context, social network interventions have been applied for mitigating HIV transmission \cite{heckathorn1999aids,rice2012mobilizing}, diabetes management \cite{vissenberg2016impact,vissenberg2017impact,vissenberg2017development} and reducing smoking \cite{pechmann2017randomised,tsoh2015social}. Social network structure is also relevant for understanding and mitigating polarisation in public sentiment concerning topics such as climate change \cite{williams2015network,kaiser2017alliance,kahan2012polarizing} and vaccine sentiment~\cite{schmidt2018polarization}. This observation has motivated the development of strategies for optimally influencing the spread of a contagion \cite{kempe2003maximizing,chen2010scalable,wang2012scalable} and equilibrium models for opinion dynamics \cite{lynn2016maximizing,lynn2017statistical,hindes2017large,lynn2018maximizing}. In this paper we model opinion dynamics using the Ising model \cite{mackay2003information,newman1999monte}. In the Ising model, the binary state of each individual is dependent on the combination of its intrinsic bias and the states of its neighbours. Ising models and their extensions are used to model phenomena including opinion dynamics \cite{castellano2009statistical,acemoglu2011opinion}, cellular signalling \cite{weber2016cellular}, neurons \cite{lynn2019physics} and ecological systems \cite{nareddy2020dynamical}. They also are an example of a Markov Random field \cite{kindermann1980markov,possolo1986estimation}, which are used in spatial statistics  \cite{cressie1992statistics,besag1974spatial} and Bolztmann machine learning \cite{tanaka1998mean-field}. Quantifying the response of Ising models to ambient fields allows us to devise public health campaigns \cite{godoy2021inference,burioni2015enhancing}, control the state of quantum systems \cite{day2019glassy,rotskoff2017geometric} and potentially devise targeted therapies for mental illness \cite{lynn2019physics}.

When modelling behaviour in large-scale social systems it is often convenient to consider the connectivity between groups of people partitioned on attributes such as age, income, ethnicity, geographic region or education level \cite{hoffmann2020inference,mcpherson2019network,godoy2021inference,butts2011spatial,blau1977macrosociological}. This approach addresses issues of privacy, reproducibility and generalizability associated with population-level social network data. Aggregation also offers significant computational advantages when considering large systems. One method to estimate social connectivity between different groups is via social surveys \cite{hoffmann2018partially-observed,godoy2021inference,mcpherson2019network}. This might be achieved via ego network surveys \cite{smith2019continued,smith2015global} or collecting aggregated relational data \cite{mccormick2015latent,breza2017using,breza2019consistently}. An alternative to direct collection of network data is to fit a statistical model to behavioural data. Given observations of some binary variable (e.g. smoker/non-smoker) it is possible to jointly infer the connectivity and impact of external covariates \cite{godoy2021inference,burioni2015enhancing,gallo2009parameter,gallo2009equilibrium,opoku2019parameter,fedele2013inverse,fedele2017inverse,peixoto2019network,schaub2019blind}. Ising models have been fit to data concerning social outcomes including voting behaviours \cite{godoy2021inference}, health screening campaigns \cite{burioni2015enhancing}, civil vs religious marriages, divorces and suicidal tendencies \cite{gallo2009parameter,gallo2009equilibrium} and educational attainment \cite{opoku2019parameter}. 

Given an estimate of the network structure we can investigate how the state of the system responds to perturbations. For Ising systems, these might include perturbations to the temperature, connectivity or external fields (see Ref. \cite{godoy2021inference} for a range of examples). In this work we focus on weak but sustained perturbations to the external fields. This type of influence has been considered recently in Refs. \cite{lynn2016maximizing,lynn2017statistical,lynn2018maximizing}. Real social systems might not respond in the same way to external perturbations, however, investigating the impact of such perturbations can provide us with plausible influence strategies and informed null models for application in practice.

In this paper we consider influencing Ising systems on networks under the following assumptions:
\begin{enumerate}[label=(\alph*)]
	\item{\textbf{Homophilous modular networks.} Social networks are well known to be homophilous with respect to attributes \cite{mcpherson2001birds,smith2014social,hoffmann2020inference}. Assuming the attributes are discrete (which is often the case as continuous variables might be binned) we can model social connectivity using a stochastic block model (SBM) \cite{karrer2011stochastic,holland1983stochastic,faust1992blockmodels}.}
	\item{\textbf{Strong ambient social fields.}  At the level of populations, an individual’s socio-demographic attributes will be highly predictive of their social outcomes \cite{godoy2021inference,opoku2019parameter,burioni2015enhancing}. In the context of Ising models, this corresponds to a dominating effect from ambient fields (with weaker fields being sufficient in the low temperature setting). We will assume that these external fields are specified by block membership. This reflects the fact that data which respects privacy and availability constraints may be aggregated.}
	\item{\textbf{Estimates of the system state.} An estimate of the system state corresponds to full or partial information about the binary state of each individual. Historically, surveys which record opinions combined with participant demographics are more prevalent than more recent data collection efforts which also include social network data.
	}
	\item{\textbf{Moderate time horizon.} Ising systems with general connectivity structure and ambient fields can have a complicated solution structure with many metastable states \cite{aspelmeier2006free-energy}. For opinion dynamics on weakly coupled modular networks the different metastable states correspond to solutions where spins within each block are mostly aligned in the positive or negative directions \cite{suchecki2009bistable-monostable,lambiotte2007coexistence,lambiotte2007majority}. When simulating Ising systems, the system will equilibrate to a metastable state, and then remain in it for an exponentially long amount of time before transitioning to another due to fluctuations \cite{hindes2017large}. We aim to identify influence strategies which are effective \emph{given knowledge of the current metastable solution.} This assumption amounts to identifying practically relevant \emph{`moderate timescale influence strategies'}, which will differ from those which are optimal in the infinite time horizon case.
	
	The influence strategies we consider are open-loop and time-independent as they do not rely on information about the system state, apart from an initial estimate of the current metastable state. Open-loop controls with pulsed fields might present a significant improvement for a given average field budget.}
	
\end{enumerate}

While these assumptions are somewhat limiting they largely correspond to the parameter regimes in which Ising influence has been studied in social networks \cite{burioni2015enhancing,opoku2019parameter,godoy2021inference}. To the authors knowledge, the problem of influencing Ising systems on modular networks in the presence of external fields has not been studied in detail. Our work illustrates how this can be achieved and identifies some of the key assumptions required.

\emph{Outline.} Sections \ref{SBM_section}, \ref{Ising_systs_sect} and \ref{IIM_definition_sect} introduce SBMs, the Ising model and the Ising influence problem. Following this, we show how a coarse-grained influence strategy can perform comparably to one which uses knowledge of the full graph in a two-block SBM (Section \ref{ising_scalable}). We show that our method is effective for a range of parameter values in coarse-grained systems where the average block degrees are sufficiently heterogeneous. In Section \ref{three_block_sbm_section} we explore how optimal influence strategies are impacted by the temperature and metastable state in the presence of ambient fields. We find that a relevant heuristic in the low field budget limit is the ability to identify blocks of nodes which are not polarised strongly in a particular direction. We then suggest some heuristic strategies for influencing Ising systems in the presence of ambient fields. In Section \ref{Pokec_influence_sect} we study the performance of coarse-grained influence strategies on a large attributed social network in which nodes can be partitioned based on attribute knowledge (age, location). We find that our coarse-grained influence strategies can achieve a macroscopic fraction ($10-50\%$) of the performance of the algorithm which uses full knowledge of network structure.

\section{Methods}

\subsection{Ising systems on networks} \label{Ising_systs_sect}

In the Ising model \cite{mackay2003information,newman1999monte}, each node $i$ is assigned a spin, $s_i \in \{-1,+1\}$ (a summary table of the symbols used in this paper can be found in Supplementary Section \ref{symbols}). In the context of social networks this might represent the opinions of individuals about a political issue or an opinion on a scientific topic such as climate change or vaccines. The spin configuration is described by the vector $\underline{s}=(s_1,s_2,...,s_N)$. A particular individual's spin is a random variable which is dependent on that of its neighbours. The spins are coupled via the graph with adjacency matrix $A$, which we will assume to be undirected with no self loops.

In the Ising model the spins tend to align so that the energy is minimized. The energy is given by:
\begin{equation} \label{ising_energy}
\mathcal{E}(\underline{s}) = -\frac{1}{2} \sum_{ij} A_{ij} s_i s_j - \sum_i g_i s_i ,
\end{equation}
where $g_i$ represents an external field acting on spin $i$. It will be appropriate to decompose $\underline{g}$ into a component $\underline{h}$ which we can control and an ambient field $\underline{b}$ which we cannot influence so that: $g_i = h_i + b_i$. The probability of the system being in a particular state, $\underline{s}$ is given by the \emph{Boltzmann distribution}:
\begin{equation} \label{boltzman_dist}
\mathbb{P}(\underline{s}) = \frac{1}{Z} e^{-\beta \mathcal{E}(\underline{s})},
\end{equation}
where $\beta$ is the inverse temperature and $Z=\sum_{\underline{s}} e^{-\beta \mathcal{E}(\underline{s})}$ is the partition function. This tells us that systems are exponentially less likely to be in states with higher energy. Evaluating this expression analytically requires a sum over $2^N$ possible values of $\underline{s}$. This is not feasible for moderately sized systems. Consequently, evaluating the distribution over states typically requires Monte Carlo simulations (see Supplementary Section \ref{Ising_monte_carlo_sect}) or analytic approximations.

The state of an Ising system can be summarised by the magnetisation:
\begin{equation}
M = \frac{1}{N} \sum_{i=1}^N \langle s_i \rangle,
\end{equation}
where $\langle \cdot \rangle$ represents an ensemble average over the dynamics. We will have $M= \pm 1$ if all of the $N$ spins are aligned in the positive or negative directions. In this study, we seek to find external fields $\underline{h}$ which maximise $M$ (taking the positive direction as a default) given some constraints. In the context of political votes, this would correspond to finding the optimal way to distribute a budget in order to maximise the vote share for a particular party.

Ising systems on networks with modular structures have been studied widely in the context of the \emph{multi-species Curie-Weiss model} \cite{fedele2013inverse,fedele2017inverse,opoku2018multipopulation} (MSCW). This model is equivalent to the Ising model on a weighted complete graph. In the MSCW model the graph is assumed to be fully connected. Ising systems on non-fully connected systems have been considered in the context of the Kernel Blau Ising model \cite{godoy2021inference}, coupled Barab\'{a}si-Albert networks \cite{suchecki2006ising,suchecki2009bistable-monostable} and SBMs \cite{lowe2019multi-group}. In what follows we will consider Ising systems on SBMs which is similar to the MSCW but with noise in the connections.

\subsection{Stochastic block models} \label{SBM_section}


Consider an undirected graph consisting of $N$ nodes, each of which belongs to one of $q$ non-intersecting subsets with fixed sizes: $N_i$ for $i \in \{1,...,q\}$. Let $\underline{t}$ be the vector describing the block membership of the individual nodes. That is, $t_i = k$ if node $i$ is in block $k$. We assume that the block assignments of the nodes are fixed. The connection probabilities of nodes in an SBM are determined by the entries of a $q \times q$ affinity matrix, $\Omega$. The elements of the symmetric adjacency matrix $A$ are Bernoulli trials with a connection probability given by:
\begin{equation}
\mathbb{P}(A_{ij}=A_{ji}=1) = \Omega_{t_i,t_j}.
\end{equation}
The \emph{expected adjacency matrix}, $\mathbb{E}(A)$, will be an $N \times N$ block matrix with elements $\Omega_{t_i,t_j}$.

Another matrix which we will use is the \emph{coupling matrix}, $\mathcal{K}$. This is a $q \times q$ matrix for which element $\mathcal{K}_{xy}$ represents the expected number of links from a randomly selected node in block $x$ to nodes in block $y$. This matrix is asymmetric in general, however, it will be symmetric when $N_x = N_y \quad \forall \quad x,y$. The entries of $\mathcal{K}$ will be constrained such that for $x \ne y$:
\begin{equation}
\mathcal{K}_{xy} = \frac{N_y}{N_x} \mathcal{K}_{yx} .
\end{equation}
The coupling matrix can be obtained from:
\begin{equation} \label{sbm_coupling_matrix}
\mathcal{K} = \Omega \mathcal{N},
\end{equation}
where $\mathcal{N} = \mathrm{diag}(N_1,N_2,...,N_q)$ is a matrix containing block sizes on the diagonal.

The expected total number of edges from block $x$ to block $y$ will be given by:
\begin{equation} \label{edges_from_con_prob}
E_{xy} = \begin{cases}
\frac{N_x (N_x - 1)}{2} \Omega_{xy} \quad \mathrm{if} \quad x = y \\
N_x N_y \Omega_{xy} \quad \mathrm{if} \quad x \ne y
\end{cases}
\end{equation}
Given network data at block-level, this formula can be inverted in order to obtain a point estimate of $\Omega_{ij}$.

\subsection{The Ising influence maximisation problem} \label{IIM_definition_sect}

Recent work in Refs. \cite{lynn2016maximizing,lynn2017statistical,lynn2018maximizing} has considered the problem of maximising the magnetisation of an Ising system using an external control field. They refer to this as the \emph{Ising Influence Maximisation} (IIM) problem. In Ref. \cite{lynn2016maximizing} they introduce the IIM and a mean-field based optimisation algorithm for solving the problem. The approximation performs well apart from close to the critical temperature $\beta_{c}$ (defined below). At high temperatures ($\frac{\beta}{\beta_c} \ll 1$) it is preferable to target high degree nodes, while at low temperatures ($\frac{\beta}{\beta_c} \gg 1$) it is better to spread the budget among low degree nodes. The mechanics of this problem are explored further in Ref. \cite{lynn2017statistical} which considers influence strategies in the limit of low field budgets.

In the Ising Influence Maximization problem, the Ising system is influenced both by a field $\underline{h}=(h_1,h_2,...,h_N)$ which we can control as well as an ambient field $\underline{b}=(b_1,b_2,...,b_N)$ which is not controlled. We consider a constraint on $\underline{h}$ of the form:
\begin{equation} \label{budget_constraint}
|\underline{h}| = \sum_{i=1}^N |h_i| \leq H, 
\end{equation}
where $H$ is the total field budget. In Ref. \cite{lynn2016maximizing} the \emph{Ising Influence Maximisation} problem is defined as follows: given an Ising system $A$, $\underline{b}$, $\beta$ and a budget $H$, find a feasible control field  $\underline{h}$, s.t $|\underline{h}| \leq H$, that maximises the magnetisation. That is, find an optimal control field $\underline{h}^*$, such that:
\begin{equation}
\underline{h}^* = \arg \max_{|\underline{h}| \leq H} M(\underline{b}+\underline{h}).
\end{equation}
The value of $M$ for a particular $\underline{h}$ can be estimated using the mean-field approximation. It is also possible to estimate it using higher order variational approximations \cite{lynn2018maximizing,opper2001advanced}. 

Under the mean-field approximation we can estimate the magnetisation of an Ising system by solving a set of self-consistency equations \cite{yedidia2001idiosyncratic,mackay2003information}:
\begin{equation} \label{ising_mf_equations}
m_i = \tanh\bigg( \beta \big( \sum_{j=1}^N A_{ij} m_j + b_i + h_i \big)  \bigg),
\end{equation}
for $i=1,...,N$. These equations can be evaluated numerically using fixed point iteration (Algorithm \ref{mf_mag_alg} in Supplementary Section \ref{Mean_Field_Ising}). We will denote the average mean-field magnetisation by $M^{MF} = \frac{1}{N} \sum_{i=1}^{N} m_i$.

For small values of $\beta$ and $\underline{h}=\underline{b}=0$, Equation. \ref{ising_mf_equations} has a single solution at $\underline{m}=0$. For larger values of $\beta$ there exist solutions with non-zero magnetisation. The point at which these non-zero solutions become stable is associated with the \emph{critical (inverse) temperature} $\beta_c$, which we use to set the temperature scale. $\beta_c$ can be identified by considering the conditions required for $\underline{m}=0$ to be the only solution to Equation \ref{ising_mf_equations} (see Supplementary Section \ref{crit_temp_discuss}).

\subsubsection{Block-level Ising influence}

For systems with block structure we can approximate the magnetisation further. In Ref. \cite{lowe2019multi-group} the authors show, for an Ising system on a two-block SBM, in the $N \rightarrow \infty$ limit, the average magnetisation of each block converges to that of the fully connected model if the graph is sufficiently dense and homophilous. We make the strong, but intuitive, assumption that these results can be extended to the case of a general number of blocks. This is not likely to be the case for all parameter regimes, however, our results indicate that this form of mean-field approximation performs well for a wide range of relevant parameter values.

For Ising systems on networks with block structure we can estimate the magnetisation at block-level by solving a coarse-grained version of Equation \ref{ising_mf_equations}:
\begin{equation} \label{mf_equations_block_level}
m_{B_x} = \tanh\bigg( \beta \big(\sum_{y=1}^q \mathcal{K}_{xy} m_{B_y} + \tilde{b}_{B_x} + \tilde{h}_{B_x} \big) \bigg),
\end{equation}
for block index $x=1,...,q$, where $\underline{m}_{B}$, $\underline{\tilde{b}}_{B}$ and $\underline{\tilde{h}}_{B}$ are, $q$-dimensional vectors of the magnetisation, ambient field, and control field at the level of blocks. We assume that all nodes in a given block experience the same ambient field. The derivation of this equation is given in Supplementary Section \ref{block_level_approx_alg}.

Given some $\underline{\tilde{h}}_{B}$, we can project the control from $q$ blocks to $N$ nodes by computing:
\begin{equation}
\underline{h}_{\mathrm{block}} = G \underline{\tilde{h}}_{B} , 
\end{equation}
where $G$ is the block membership matrix for which $G_{ij}=1$ if node $i$ is in block $j$ and $G_{ij}=0$ otherwise.

\subsubsection{Ising influence in the low $H$ limit. }

Apart from in Section \ref{ising_scalable}, we will consider values $H$ which are sufficiently small to induce changes in $M$ of only a few percent. Improvements of this magnitude constitute substantial progress in social applications. We can characterise the low field budget solutions to the IIM problem by computing the susceptibility vector $\nabla_{\underline{h}} M^{MF}|_{\underline{h}=0}$ \cite{lynn2017statistical} (see Supplementary Section \ref{PGA_section}).

In the following we will compare the node-specific (or full) influence strategy $\underline{h}_{\mathrm{full}}$ with the coarse-grained approximation $\underline{h}_{\mathrm{block}}$. In Section \ref{ising_scalable} we go beyond the low $H$ limit and must therefore use a projected gradient ascent (Algorithm \ref{mf_iim_alg}) approach in order to estimate $\underline{h}_{\mathrm{full}}$ and $\underline{h}_{\mathrm{block}}$. In Sections \ref{three_block_sbm_section} and \ref{Pokec_influence_sect} we assume $H$ is sufficiently small that we can set $\underline{h}_{\mathrm{full}}=\nabla_{\underline{h}} M^{MF}|_{\underline{h}=0}$ and $\underline{h}_{\mathrm{block}}=\nabla_{\underline{\tilde{h}}_{B}} M^{MF}|_{\underline{\tilde{h}}_{B}=0}$ without a significant impact on the performance.

\section{Results}

\subsection{A scalable approach for influencing Ising models on networks with block structure} \label{ising_scalable}

\begin{figure*}
	\centering
	
	\hspace*{1.2em}
	\subfloat[]{\label{control_on_two_block_graph_plot}\includegraphics[width=.64\textwidth]{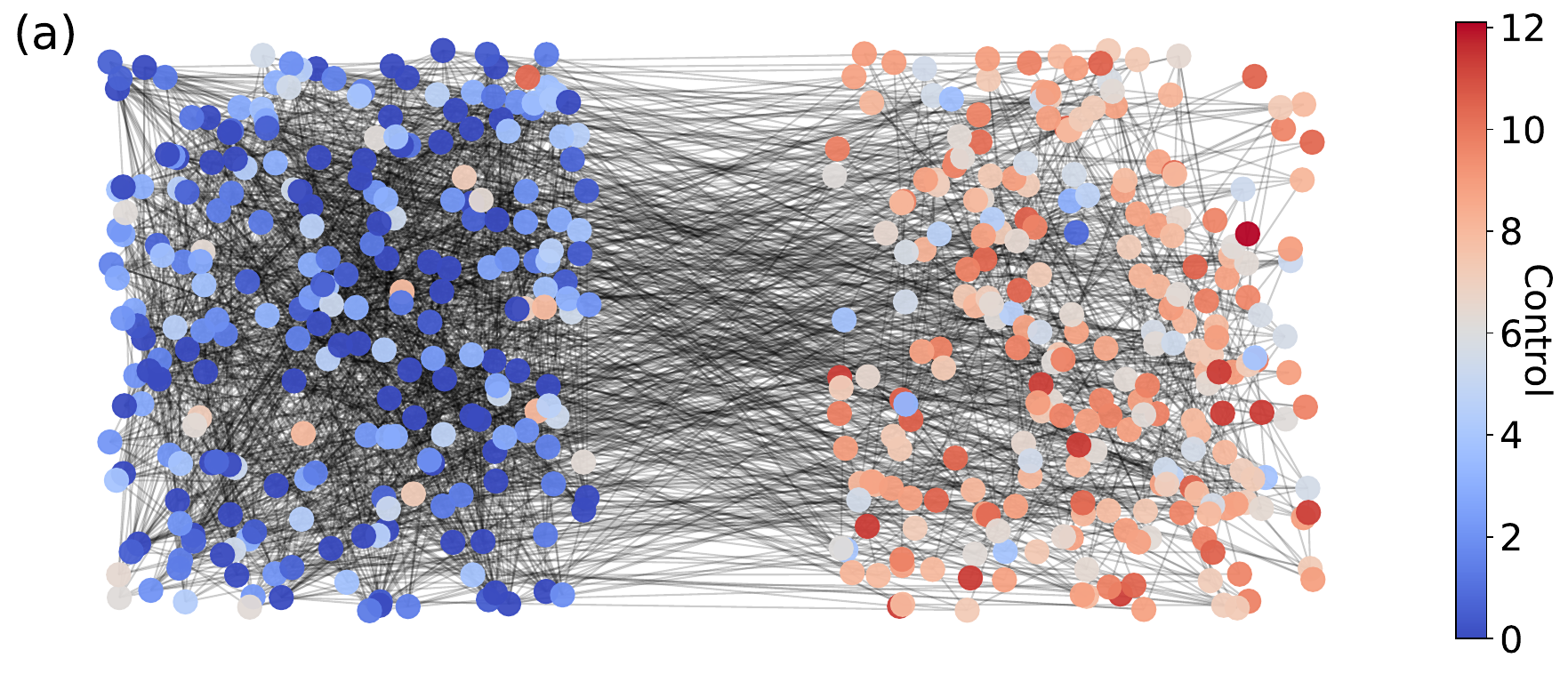}}
	\subfloat[]{\label{two_block_control_hist}\includegraphics[width=.32\textwidth]{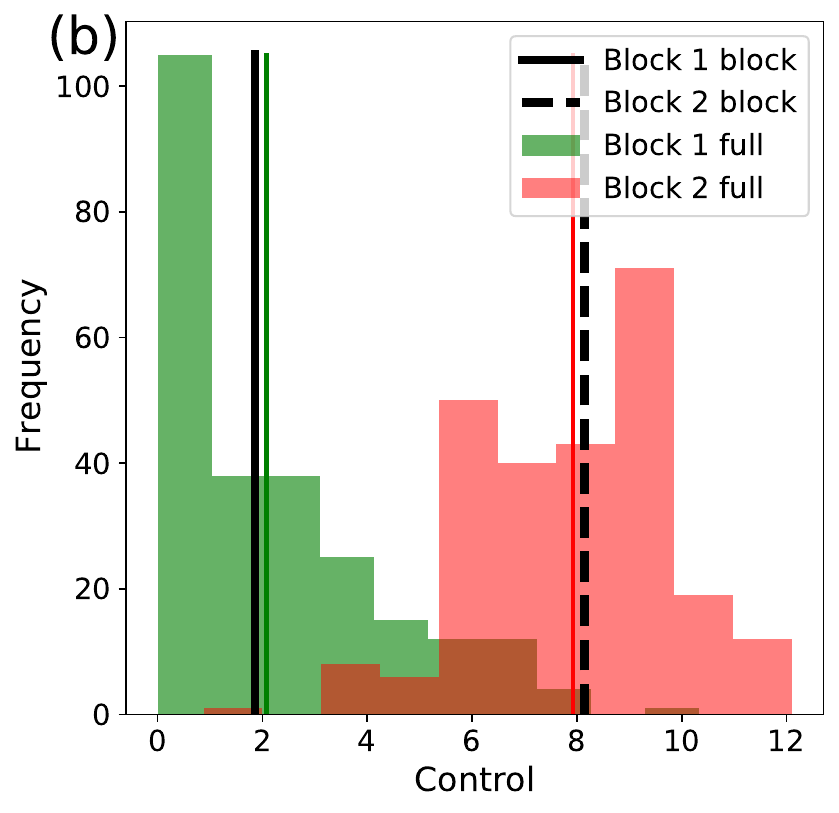}}
	
	\vspace*{-3.0em}
	\subfloat[]{\label{markup_bet05}\includegraphics[width=.32\textwidth]{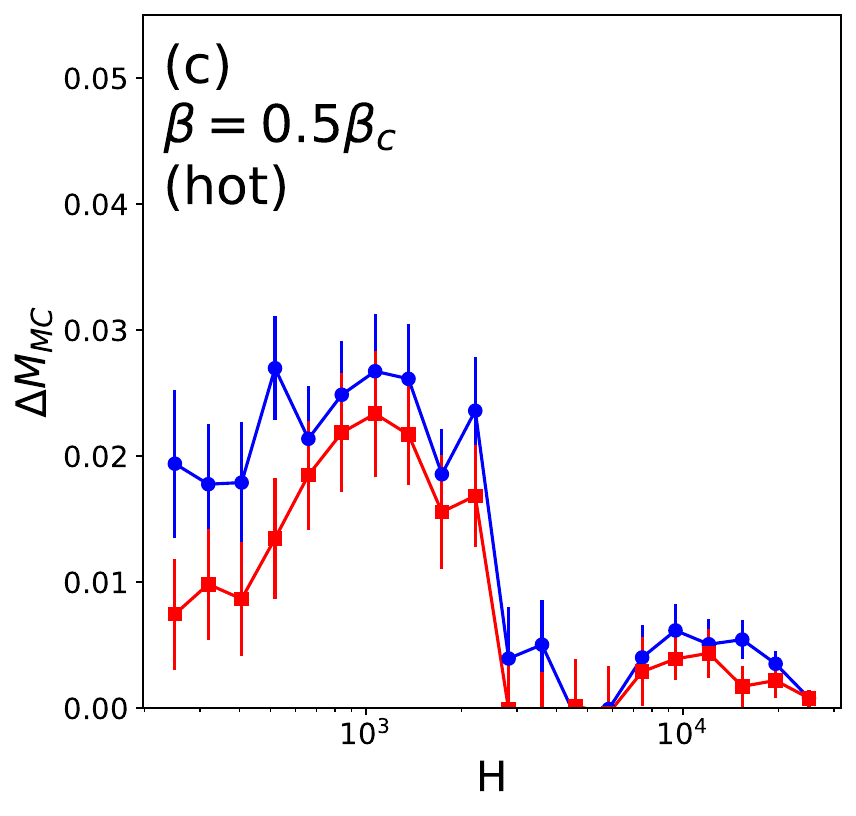}}
	\hspace*{0.6em}
	\subfloat[]{\label{markup_bet12}\includegraphics[width=.32\textwidth]{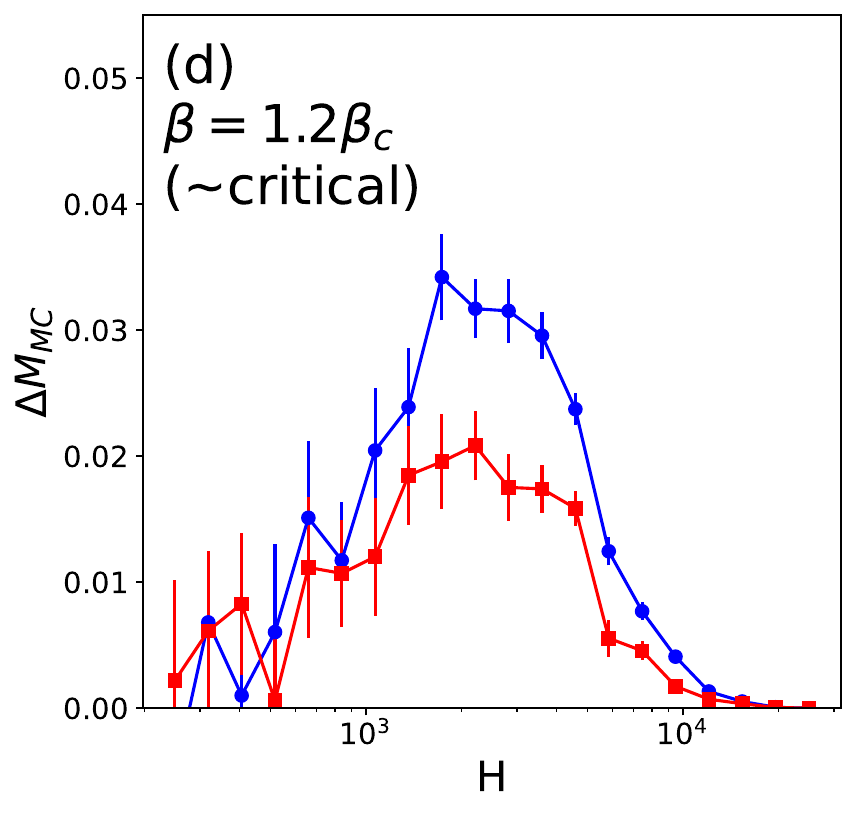}}
	\hspace*{0.6em}
	\subfloat[]{\label{markup_bet15}\includegraphics[width=.32\textwidth]{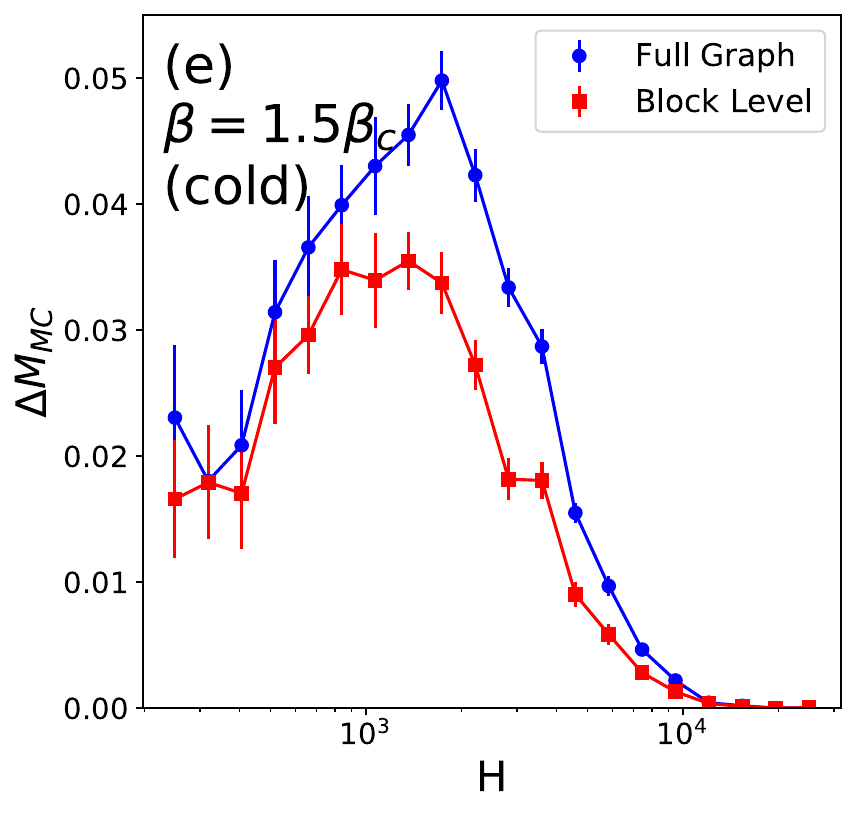}}
	\vspace*{-3.0em}
	
	\subfloat[]{\label{mag_as_H_fig1}\includegraphics[width=.32\textwidth]{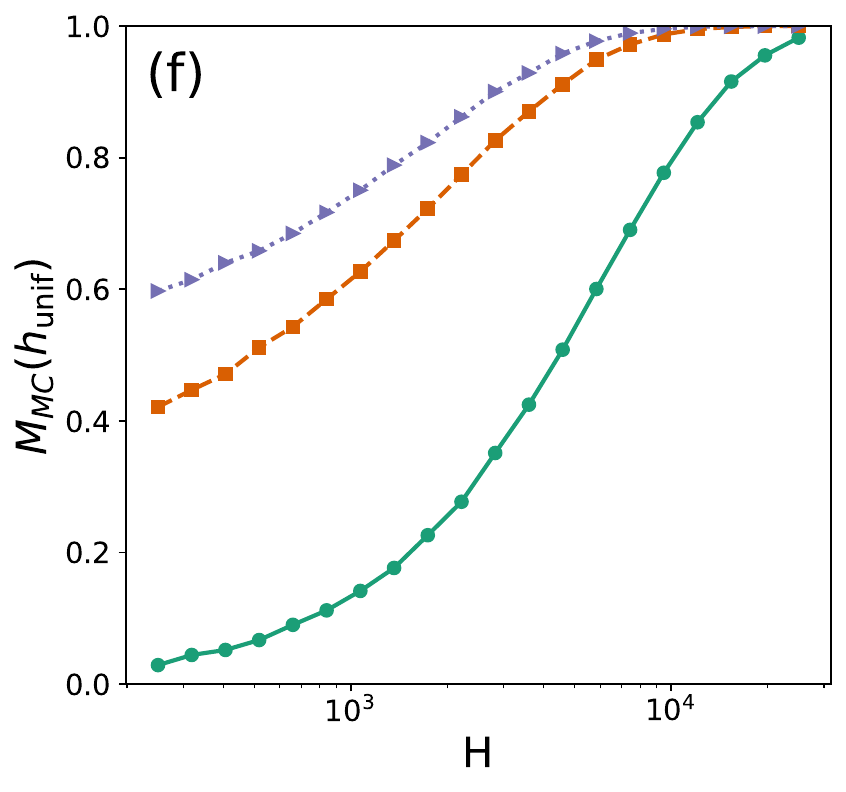}}
	\subfloat[]{\label{frac_markup_fig1}\includegraphics[width=.3345\textwidth]{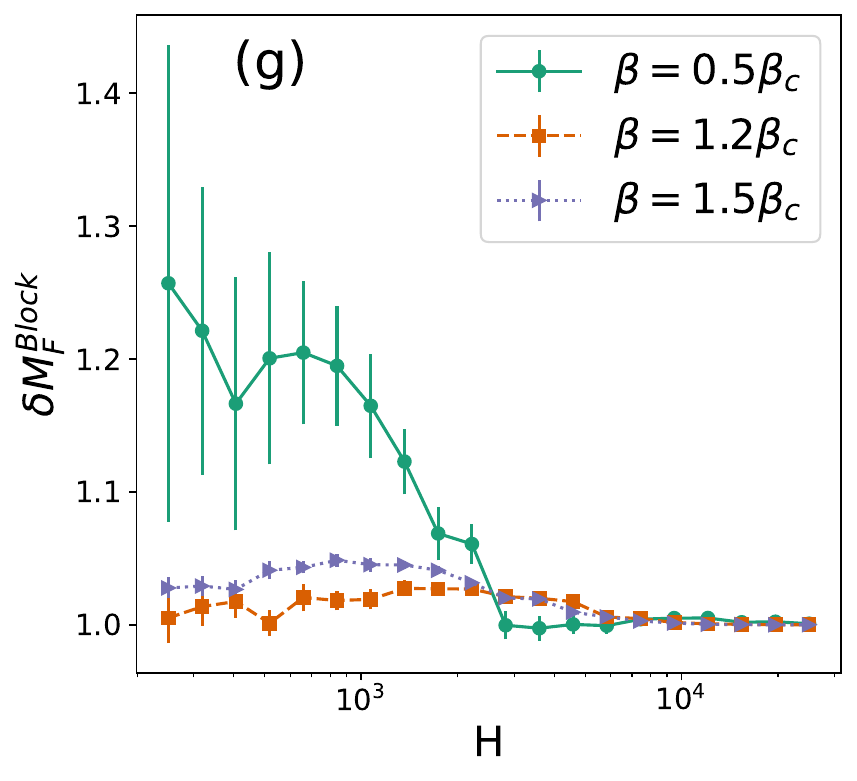}}
	\vspace*{-2.0em}
	\caption{\textbf{The block-level control can perform comparably to a control which uses the whole graph structure on a two-block SBM}. (a) Plot showing the SBM defined in text with nodes coloured according to the value of the full graph control. Nodes in block 1 (LHS) have an average degree of 12.5 while those in block 2 (RHS) have a smaller average degree of 5. (b) Histogram showing the distribution of $\underline{h}_{\mathrm{full}}$ values for the two blocks for $\beta =  \frac{3}{2} \beta_c$, $H=10$. The average control applied to the second, lower degree, block is significantly smaller than that applied to the first block. We evaluated the controls on the same SBM network drawn from the ensemble with coupling matrix in Equation \ref{two_block_coupling_matrix} for a range of $\beta$ and $H$ values. (c), (d), (e) Illustrate how $\Delta M(\protect\underline{h}_{\mathrm{full}})$ (blue circles) and $ \Delta M(\protect\underline{h}_{\mathrm{block}})$ (red squares) behave as a function of $H$ for $\beta = 0.5 \beta_c$ (c), $1.2 \beta_c$ (d) and $ 1.5 \beta_c$ (e). (f) and (g) show the behaviour of $M_{MC}(\underline{h}_{\mathrm{unif}})$ and $\delta M_{F}^{Block}$ respectively for different values of $\beta$. Error bars indicate the standard error on the mean from 15 Monte Carlo simulations. Parameters used for the numerical simulations are given in Supplementary Section \ref{Parameters_two_block}.}
	\label{two_block_markups}
\end{figure*}

In this section we compare the structure of the solutions obtained by the full graph and the block-level influence strategies on a two-block SBM. We then show that the magnetisation obtained using the block-level strategy can be comparable to that obtained by using the full graph strategy on the same SBM.

In the absence of external fields, the leading order effect specifying the optimal nodes to target is typically their degree \cite{lynn2016maximizing,lynn2017statistical}. Consequently, we will focus on a system for which the average degrees are heterogeneous. If the average degrees of blocks are homogeneous we expect little impact from coarse-grained influence strategies (this is confirmed in the zero ambient field setting in Supplementary Section \ref{degree_hetero_impact}).

We will consider an Ising system on the two-block SBM with the coupling matrix:
\begin{equation}  \label{two_block_coupling_matrix}
\mathcal{K} = \begin{pmatrix}
10 & 2.5 \\
2.5 & 2.5
\end{pmatrix}.
\end{equation}
This system consists of a strongly self-coupled block (block 1) coupled to a weakly
self-coupled one (block 2). We set the block sizes to be equal with $N_1=N_2=250$. A draw from this ensemble is shown in Figure \ref{control_on_two_block_graph_plot}.

We evaluate the performance of $\underline{h}_{\mathrm{full}}$ and $\underline{h}_{\mathrm{block}}$ using Monte Carlo simulations and explore how the performance is impacted by the parameters $H$ and $\beta$. Let $M_{MC}(\underline{h})$ be the average magnetisation of the system evaluated using Monte Carlo simulations. Define the \emph{magnetisation markup} for control $\underline{h}$ to be:
\begin{equation} \label{mag_markup_def}
\Delta M_{MC}(\underline{h}) = M_{MC}(\underline{h}) - M_{MC}(\underline{h}_{\mathrm{unif}}) , 
\end{equation}
where $\underline{h}_{\mathrm{unif}}$ represents the uniform baseline strategy for which $h_{\mathrm{unif},i}=\frac{H}{N}$.

We compute $\underline{h}_{\mathrm{full}}$ and $\underline{h}_{\mathrm{block}}$ for a range of values of $H$. We consider 3 different values of $\beta_c = \frac{1}{\lambda_1}$ where $\lambda_1$ denotes the largest eigenvalue of the adjacency matrix, $A$, of the sampled SBM. The parameters used in the projected gradient ascent algorithm are provided in Supplementary Section \ref{Parameters_two_block}.

The average magnetisations for controls were estimated by running Markov-Chain Monte Carlo with the Metropolis dynamics described in Supplementary Section \ref{Ising_monte_carlo_sect}. Simulations were initialised with all spins pointing in the positive direction ($\underline{s}=(1,1,...,1)$) with the aim of identifying the most positive metastable solution. For each value of the field budget we determine the magnetisation by taking the average value obtained from 15 simulations with a burn-in time of $T_{\mathrm{burn}} = 2 \times 10^4$ and run time of $T=10^4$ steps. 

Figures \ref{markup_bet05}, \ref{markup_bet12} and \ref{markup_bet15} show $\Delta M_{MC}(\underline{h})$ as a function of $H$ for $\beta = 0.5 \beta_c , 1.2 \beta_c$ and $ 1.5 \beta_c$ respectively. $\Delta M_{MC}(\underline{h}_{\mathrm{block}})$ lies close to $\Delta M_{MC}(\underline{h}_{\mathrm{full}})$ for a range of values of $H$ and $\beta$. This indicates the presence of scenarios in which the block-level strategy can be used to effectively Ising influence Ising systems on SBMs. We would expect $\Delta M_{MC}(\underline{h}_{\mathrm{block}})$ to be much smaller in the case where the two blocks have similar average degrees.

We also compute the quantity $\delta M_{F}^{Block}= \frac{M_{MC}(\underline{h}_{\mathrm{block}})}{M_{MC}(\underline{h}_{\mathrm{unif}})}$ (Figure \ref{frac_markup_fig1}) which indicates the fractional increase in the magnetisation relative to the uniform baseline. $\delta M_{F}^{Block}$ is largest for $\beta=0.5 \beta_c$ for which we see a relative increase in the magnetisation of at least $20 \%$. For larger $\beta$ we observe a smaller increase of up to $\approx5\%$. The larger value of $\delta M_{F}^{Block}$ in the former case coincides with the case when the $M_{MC}(\underline{h}_{\mathrm{unif}})$ is small (Figure \ref{mag_as_H_fig1}). 

\subsection{Ising influence in the presence of ambient fields} \label{three_block_sbm_section}

In this section we explore the  influence strategies which arise in Ising systems with ambient fields. We first consider the Ising influence problem on a simple model of a polarised society. The insights drawn from this analysis provide us with some more informed baseline control strategies to be applied to Ising systems with ambient fields.

\begin{figure*}
	\centering
	
	\subfloat[]{\label{three_blocks_sbm_w_back}\includegraphics[width=.5\textwidth]{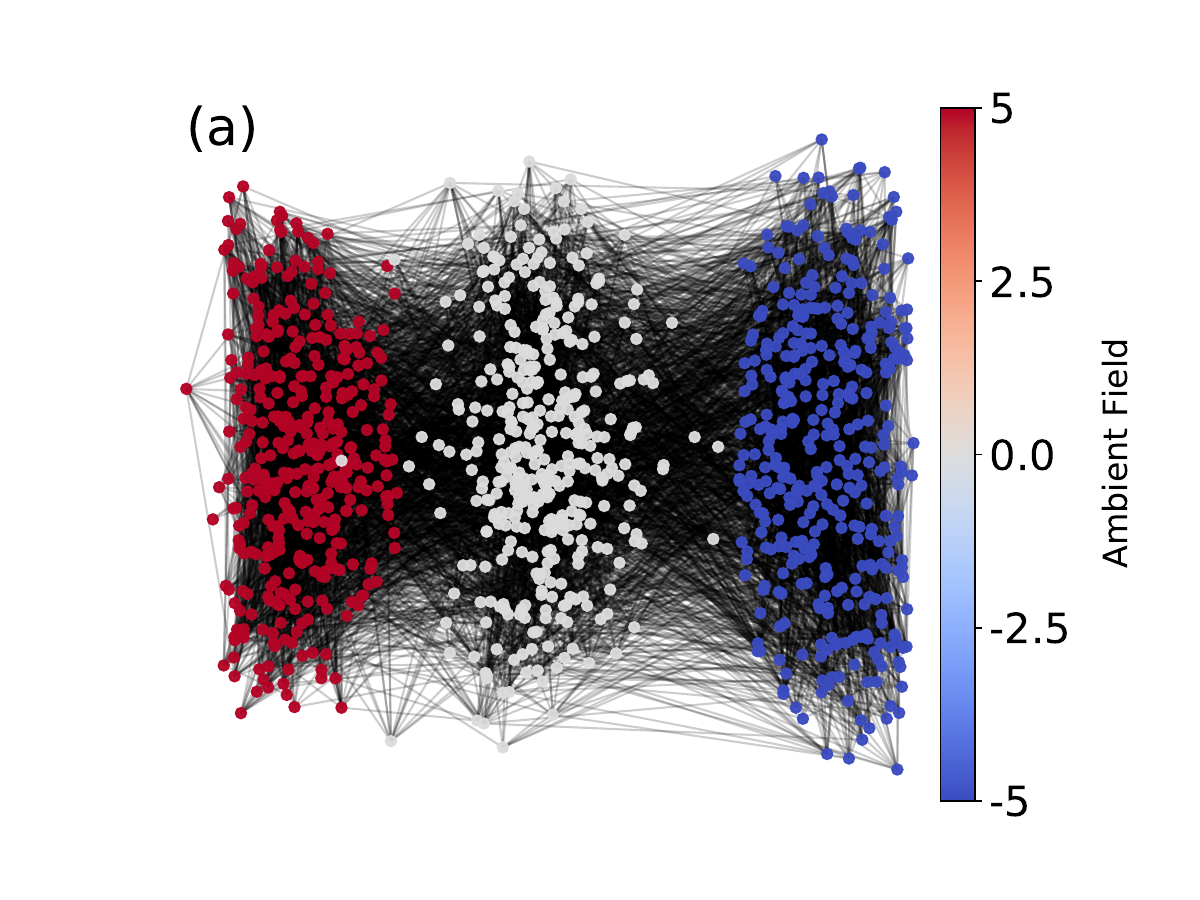}}
	\subfloat[]{\label{three_block_phase}\includegraphics[width=.5\textwidth]{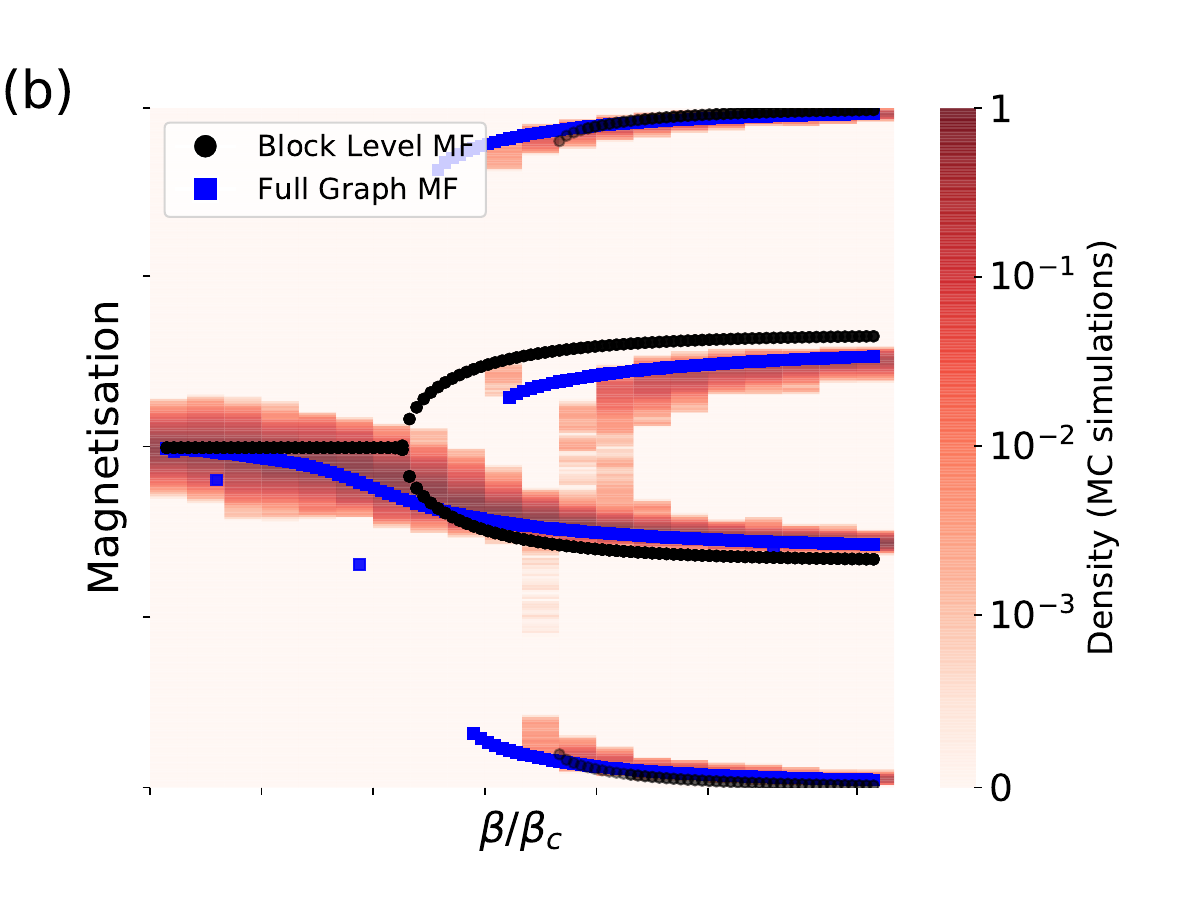}}
	
	\vspace*{-2.0em}
	\subfloat[]{\label{sus_on_phase_1}\includegraphics[width=.22\textwidth]{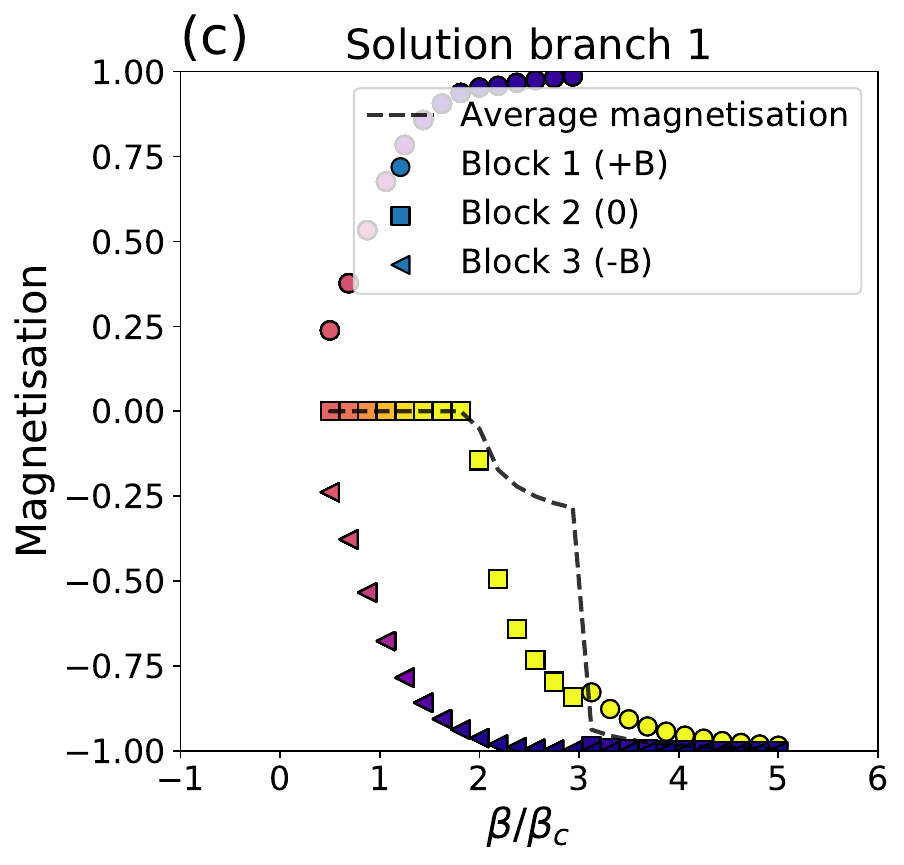}}
	\subfloat[]{\label{sus_on_phase_2}\includegraphics[width=.22\textwidth]{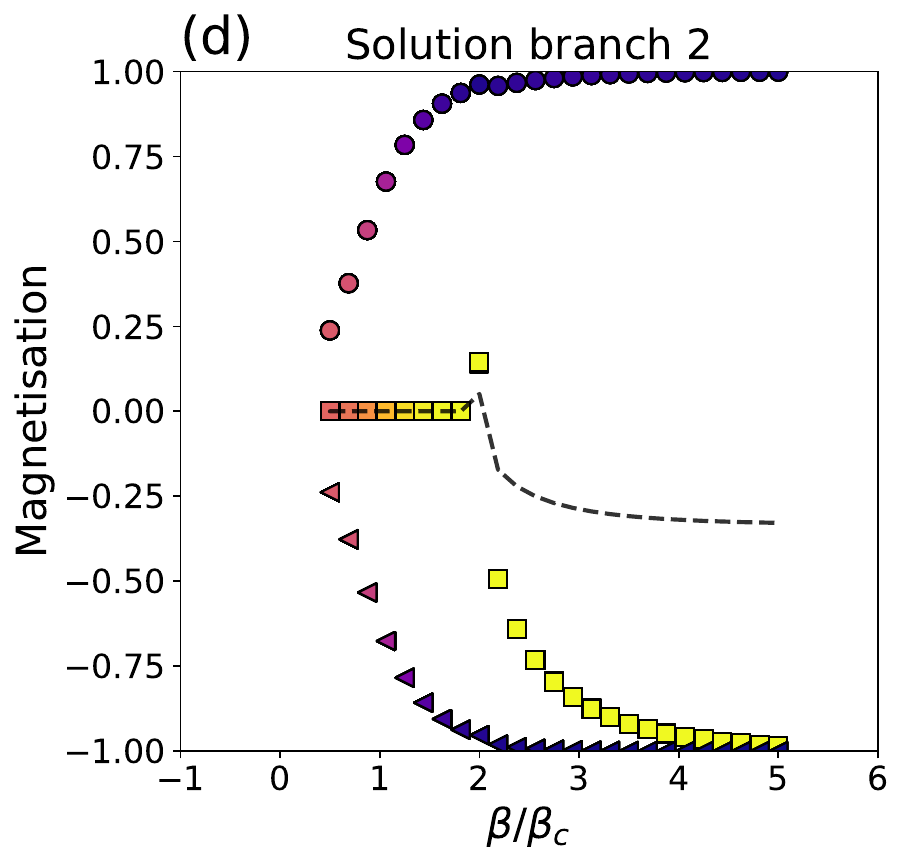}}
	\subfloat[]{\label{sus_on_phase_3}\includegraphics[width=.22\textwidth]{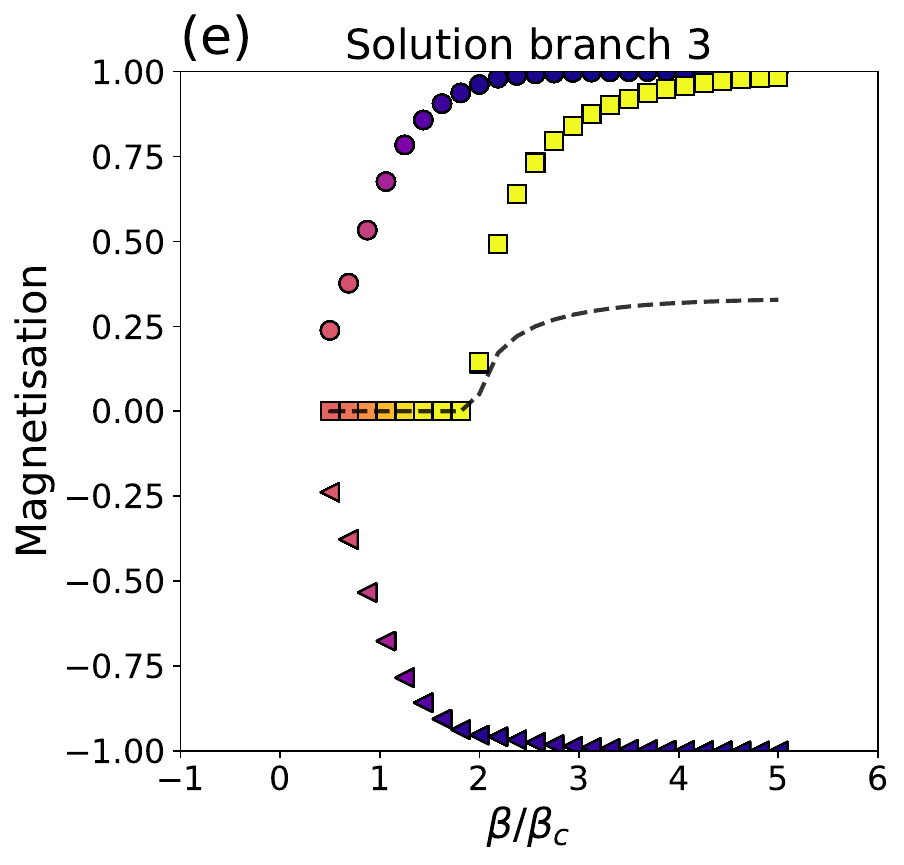}}
	\subfloat[]{\label{sus_on_phase_4}\includegraphics[width=.275\textwidth]{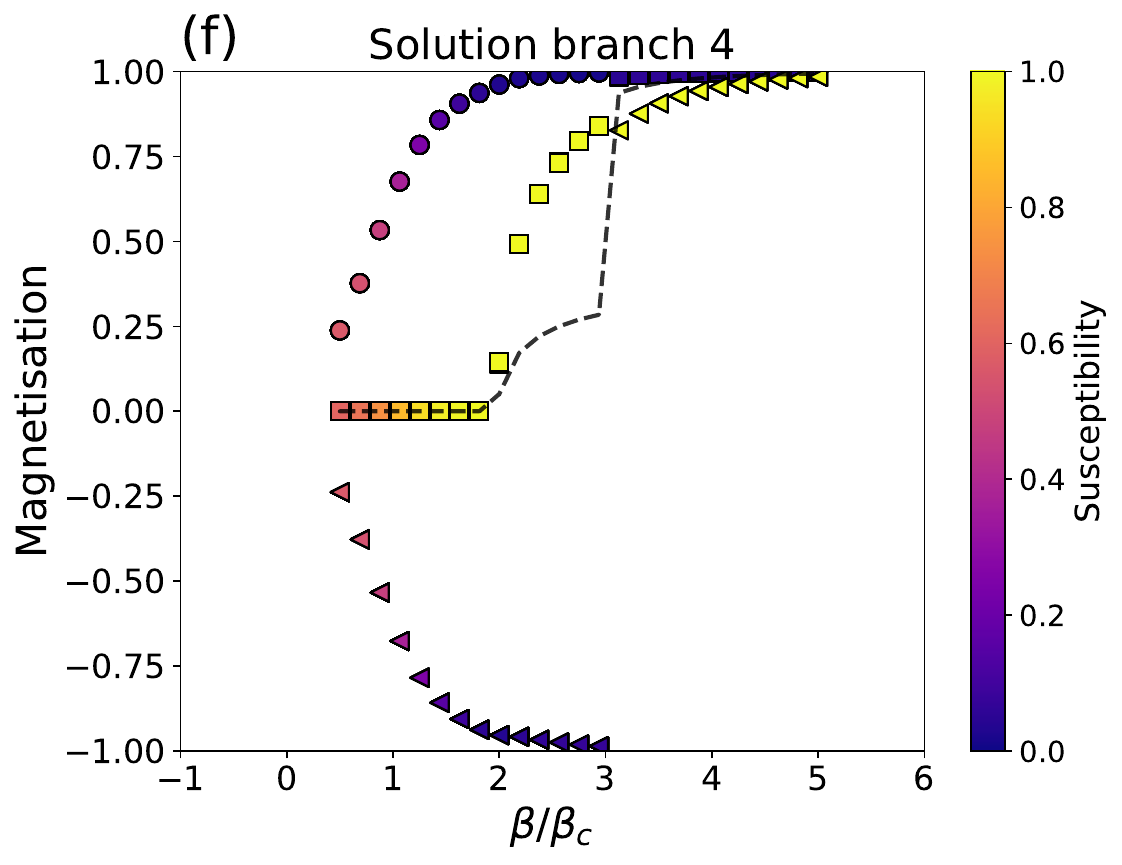}}
	\vspace*{-2.0em}
	\subfloat[]{\label{blocksus_1}\includegraphics[width=.22\textwidth]{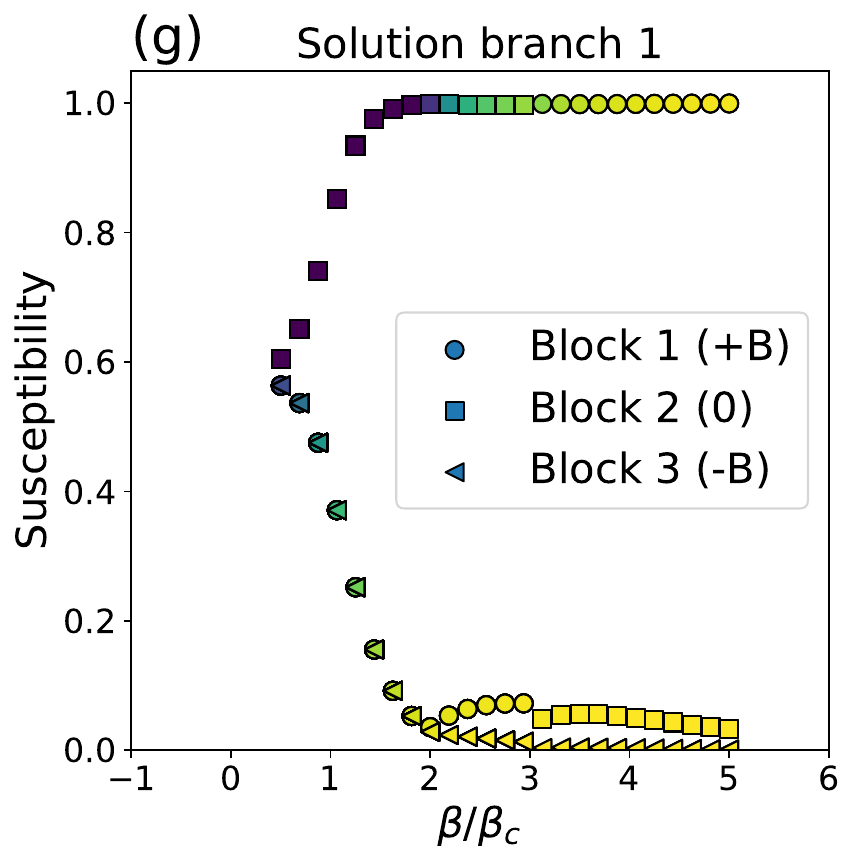}}
	\subfloat[]{\label{blocksus_2}\includegraphics[width=.22\textwidth]{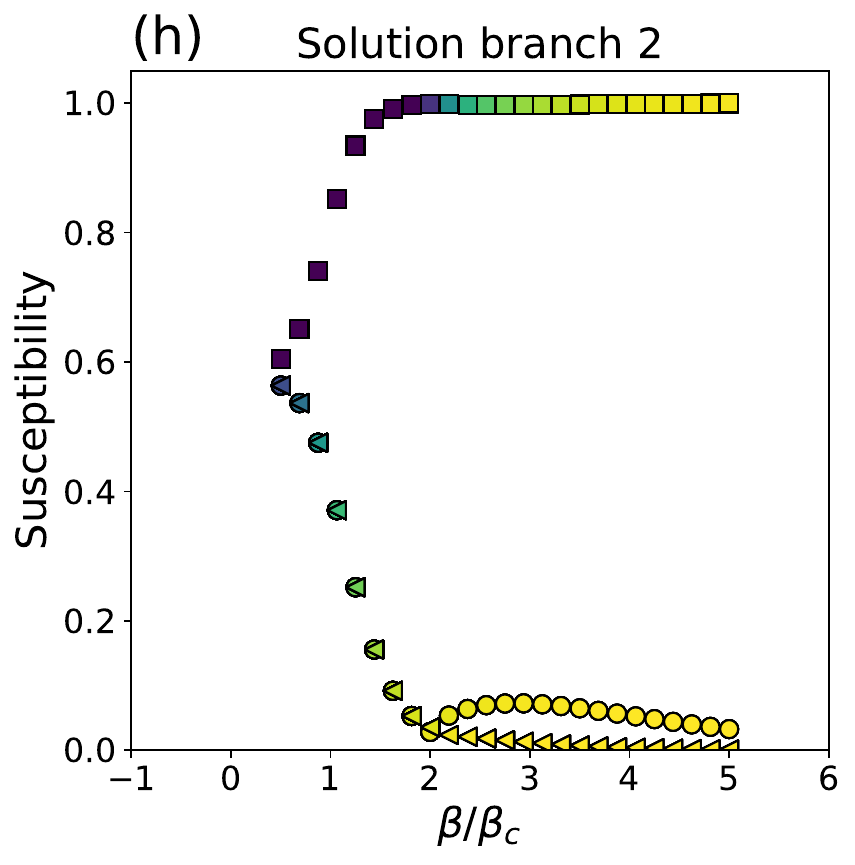}}
	\subfloat[]{\label{blocksus_3}\includegraphics[width=.22\textwidth]{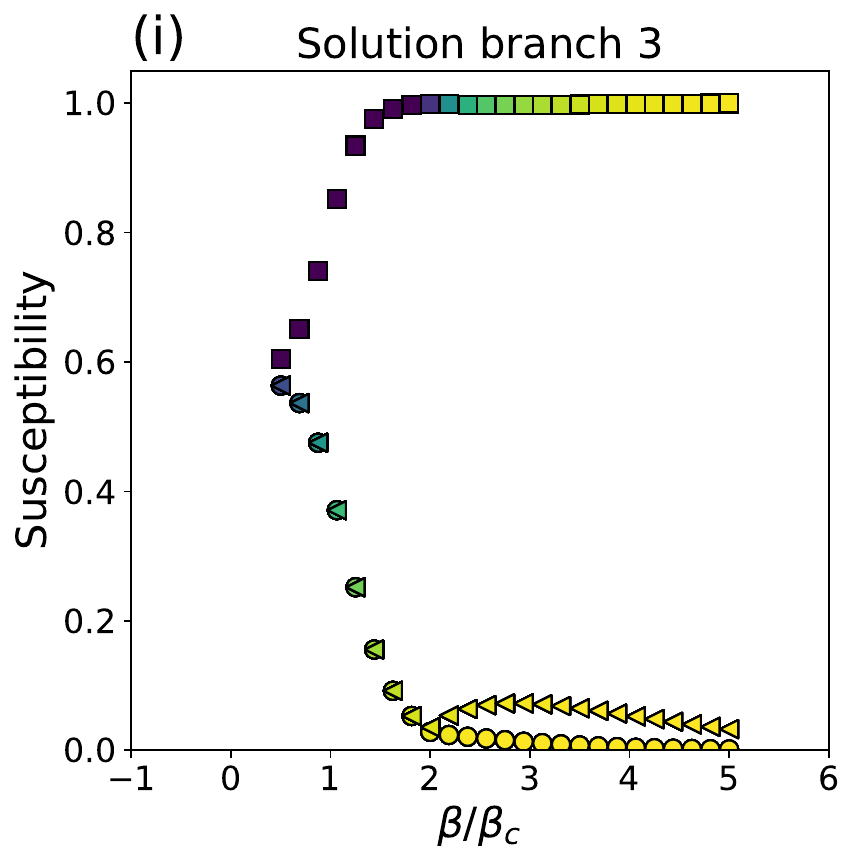}}
	\subfloat[]{\label{blocksus_4}\includegraphics[width=.275\textwidth]{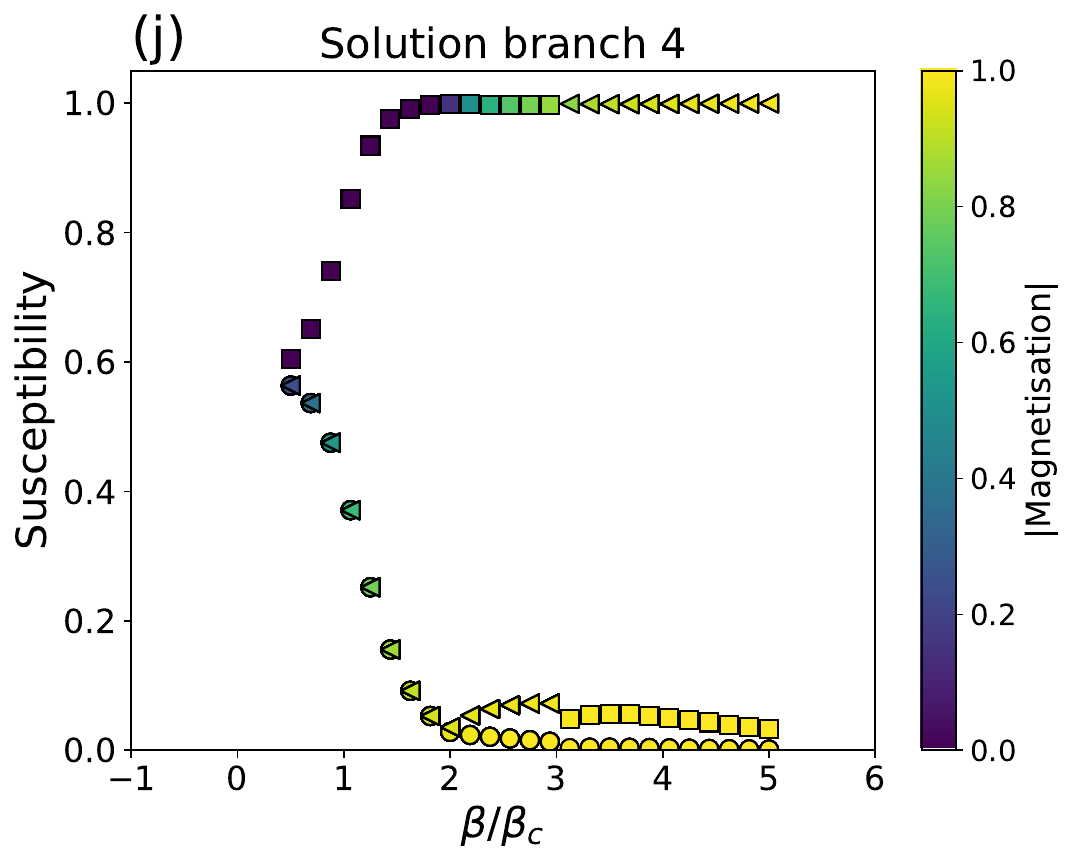}}
	
	\vspace*{-2.0em}
	\caption{\textbf{The optimal low $H$ block-level IIM strategy involves targetting the block with the lowest absolute magnetisation.} (a) SBM drawn from the ensemble with the coupling matrix specified in Equation \ref{three_block_coupling} for $N=1200$. The ambient fields experienced by the nodes are indicated on the colour bar. (b) Plot showing the different solutions for the magnetisation of the three-block system as a function of $\beta$. The plot shows the different solutions identified using the block-level mean-field approximation (black circles) and the full graph mean-field approximation (blue squares). These are overlaid onto a heatmap showing the probability distribution of magnetisation values obtained using Monte Carlo simulations. For each value of $\beta$ we sampled from two Monte Carlo chains with a burn-in time of $1.2 \times 10^5$ and run time of $10^4$ steps for each of 25 random initial conditions (see Supplementary Section \ref{three_block_phase_diag}). The histograms (for a given $\beta$) are normalised by the maximum count so that the different metastable states are easier to identify. The value of $\beta_c$ was computed from the spectral radius of the SBM. Sub-figures (c)-(f) show the behaviour of the magnetisation at the level of blocks as a function of $\beta$ computed using the block-level mean-field approximation for the different metastable solutions. Scatter points are coloured according to the component of the magnetisation gradient $\nabla_{\underline{h}} M^{MF}|_{\underline{h}=0}$ associated with each block. Dashed lines indicate the average magnetisation of the metastable solution. These metastable solutions correspond to the different branches shown in Sub-figure (b). Sub-figures (g)-(j) show the components of $\nabla_{\underline{h}} M^{MF}|_{\underline{h}=0}$ coloured according to the absolute value of the magnetisation of each block. Solutions corresponding to the different metastable states are obtained by initialising Algorithm \ref{mf_mag_alg} with different uniform random initial conditions in $[-1,1]^3$. $\nabla_{\underline{h}} M^{MF}|_{\underline{h}=0}$ is computed by numerically solving Equation \ref{linear_eq_for_sus} in each instance.}
	\label{three_block_phase_diagram}
\end{figure*}

We will consider an SBM consisting of 3 blocks arranged in a chain with the coupling matrix:
\begin{equation} \label{three_block_coupling}
\mathcal{K} = \begin{pmatrix}
10 & 2.5 & 0 \\
2.5 & 7.5 & 2.5 \\
0 & 2.5 & 10
\end{pmatrix},
\end{equation}
with block sizes of: $N_1=N_2=N_3=400$. We set the block-level ambient field vector to:
\begin{equation}
\underline{\tilde{b}}_{B} = (+B,0,-B),
\end{equation}
where the parameter $B$ controls the magnitude of the field applied. This model consists of two communities with opposing `external bias' which can only communicate via an intermediate group with no external bias. The coupling structure here shares similarity with that between three different age groups obtained by fitting an Ising model to health screening data in Ref. \cite{burioni2015enhancing} and is consonant with the geometric kernel inferred in \cite{godoy2021inference}. Figure \ref{three_blocks_sbm_w_back} shows a typical draw of an SBM from this system. The average degrees of the nodes in the SBM with the coupling matrix in Equation \ref{three_block_coupling} will be equal. This homogeneity in the node degrees allows us to focus on the effects of the ambient field on the optimal influence strategy.

Before studying the influence problem on this network, we first enumerate the values which $M$ can take for different $\beta$ values in the absence of a control field. We compute the phase diagram of the system using two different mean-field approximations (full and block-level) and Monte Carlo simulations. Different metastable solutions can be identified using different initial conditions for the fixed point iterations and Monte Carlo simulations. We identify these different metastable solutions using a sampling scheme which assumes that blocks have the same fraction of aligned spins (see methods in Appendix \ref{three_block_phase_diag}).

The phase diagram for the three block system is shown in Figure \ref{three_block_phase}. The phase diagram consists of three different regimes containing one, two and four metastable states. The results shown in Figure \ref{three_block_phase} demonstrate that the block-level mean-field approximation is able to reproduce the shape of the phase diagram obtained using Monte Carlo simulations and the full graph mean-field approximation. The main discrepancies between the different methods occur close to the two transition points. The lack of symmetry in the full mean-field approximation and Monte Carlo simulations is an artefact of the particular SBM draw which we expect will disappear under ensemble averaging (see Supplementary Section \ref{ens_var_phase_diag}).

\subsubsection{Low $H$ solutions to the IIM problem with ambient fields} \label{three_block_structure}


We characterise the low $H$ solutions to the IIM problem by computing the magnetisation and component of $\nabla_{\underline{h}} M^{MF}|_{\underline{h}=0}$ associated with each block for each of the metastable solution branches identified in Figure \ref{three_block_phase}. Figures \ref{sus_on_phase_1}-\ref{sus_on_phase_4} show how the magnetisation at the level of individual blocks varies as a function of $\beta$. We colour the blocks according to the components of $\nabla_{\underline{h}} M^{MF}|_{\underline{h}=0}$. Figures \ref{blocksus_1}-\ref{blocksus_4} illustrate how the susceptibility of each block varies as a function of $\beta$ for each solution branch. In the noisy or high temperature (low $\beta$) setting, the susceptibility of all three blocks approaches the same value. This suggests it is optimal to spread the budget uniformly, which is intuitive as the blocks have the same average degree. For large $\beta$, the optimal low $H$ block-level influence strategy is to focus all of the budget on one of the three blocks. In all cases this is the block which has the lowest absolute magnetisation. In sub-figures \ref{blocksus_1} and \ref{blocksus_4} we observe an abrupt transition in the block with the highest susceptibility. This corresponds to the formation of new solution branches where all three blocks are aligned in the same direction.

In this example, knowledge of the system parameters ($\beta$, $\mathcal{K}$ and $\underline{\tilde{b}}$) alone are not sufficient to identify the optimal medium timescale influence strategy. Instead, it is necessary to be able to identify the least polarised set of nodes. This mirrors the tactic of targetting sub-populations with no strong bias either way known as `swing voters' in political campaigns. This situation also has similarities to public health efforts to target the vaccine hesitant \cite{butler2015diagnosing}.

\begin{table*}[]
	\centering
	\resizebox{0.8\textwidth}{!}{%
		\begin{tabular}{|l|l|l|l|}
			\hline
			Influence strategy  & Definition                                                                                                                                                                                                                                                                                                                                                                   & Information required                                                 & Parameters                        \\ \hline
			Full graph          & \begin{minipage}{110mm}The \emph{normalised} magnetisation gradient $\nabla_{\underline{h}} M^{MF}|_{\underline{h}=0}$ is obtained by numerically solving Equation \ref{linear_eq_for_sus} in the Supplement. In Section \ref{ising_scalable} we use the projected gradient ascent approach presented in Algorithm \ref{mf_iim_alg}. \end{minipage}                          & $A$, $\underline{b}$, $\beta$, $\underline{m}^0$                     & $\epsilon$, $a$, $\gamma$, $\tau$ \\ \hline
			Block-level         & Identical to the above but computed for at the coarse grained level (Equation \ref{mf_equations_block_level}).                                                                                                                                                                                                                                                               & $\mathcal{K}$, $\tilde{\underline{b}}$, $\beta$, $\underline{m}_B^0$ & $\epsilon$, $a$, $\gamma$, $\tau$ \\ \hline
			Uniform             & $\underline{h}_{\mathrm{unif}}=(\frac{H}{N},\frac{H}{N},...,\frac{H}{N})$                                                                                                                                                                                                                                                                                                    & N/A                                                                  & N/A                               \\ \hline
			Hesitant targetting & \begin{minipage}{110mm}Define the set of hesitant nodes to be: $\mathcal{H}(\psi) = \{ i : |b_i| < \psi \}$, where $\psi$ is a constant. This strategy spreads influence equally among this set of nodes so that: $ h_{HT,i} = \frac{H}{|\mathcal{H}(\psi)| } \quad \mathrm{if} \quad i \in \mathcal{H}(\psi)$ and  $ h_{HT,i} = 0 \quad \mathrm{otherwise} $ \end{minipage} & $\underline{b}$ (or $\tilde{\underline{b}}$)                         & $\psi$                            \\ \hline
			Negative canceller  & \begin{minipage}{110mm}$\underline{h}_{NC} = \frac{H}{| \underline{h}^{-} |}  \underline{h}^{-}$, where $\underline{h}^{-}$ is the vector with elements: $h^{-}_i =-b_i \quad$ if $\quad b_i < 0$ and $h^{-}_i =0 \quad$ otherwise \end{minipage}                                                                                                                            & $\underline{b}$ (or $\tilde{\underline{b}}$)                         & N/A                               \\ \hline
			Survey-snapshot     & See Section \ref{Strategy_definition}.                                                                                                                                                                                                                                                                                                                                       & $\underline{\hat{M}}_B$                                              & $t$                               \\ \hline
	\end{tabular}}
	\caption{Table comparing the different Ising influence strategies and null models introduced in text. Symbols used in the manuscript are defined in Supplementary Table \ref{symbols_table}. For the full graph and block-level IIM strategies $\underline{m}^0$ and $\underline{m}_B^0$ can be any initial condition which converges to the desired metastable solution under the dynamics in Algorithm \ref{mf_mag_alg}. Setting $\underline{m}^0$ and $\underline{m}_B^0$ requires knowledge of the metastable state which the system is assumed to be in. For main text Figures \ref{two_block_markups} and \ref{pokec_w_fields_magcontrols_both} we select the metastable solutions with the most positive and most negative magnetisations respectively. These can be reached by initialising at $(1,1,...,1)$ and $(-1,-1,...,-1)$ respectively. When implementing the Survey-snapshot strategy in this paper we use Monte Carlo simulations in order to estimate the magnetisation at the level of blocks $\underline{\hat{M}}_B$. However, in practice we would aim to set $\underline{m}^0$, $\underline{m}_B^0$ and $\underline{\hat{M}}_B$ from empirical survey data (e.g. as in Refs. \cite{godoy2021inference} and \cite{burioni2015enhancing}).}
	\label{strat_summary}
\end{table*}

\subsubsection{Heuristic influence strategies and null models} \label{Strategy_definition}

The above results suggest that knowledge of the magnetisation of groups of nodes is essential for influencing Ising systems. In practice, we might only be able to obtain relatively coarse-grained information about the system state. Thus, to explore how useful this information might be in practice, we propose a heuristic strategy for influencing Ising systems which relies on point estimates of the magnetisation at the level of blocks. We will assume that we can only access a snapshot of the system at a particular time rather than relying on an aggregation of multiple observations. A more robust estimate of the susceptibility of the system would require multiple observations or knowledge of the connectivity between the blocks (which we assume is not available in the case of this null model).

In analogy to a social survey taken at a particular point in time, we will use a sample of node states to estimate the magnetisation. We assume that it is possible to estimate the magnetisation based on the spins of all of the nodes, however, in Supplementary Section \ref{snapshot_convergence} we provide evidence that this strategy performs well even when a small fraction of nodes from each block are sampled. This analysis assumes that a random sample is equivalent to a representative sample of the population, which is not necessarily the case in practice.

The strategy is motivated by the fact that differentiating Equation \ref{ising_mf_equations} with respect to $h_i$ it is possible to show that the mean-field susceptibility of a node to its own field is equivalent to $\beta(1-m_i^2)$. Let $\hat{M}_{B_x}(t)$ be an estimate of the magnetisation of block $x$ at timestep $t$. Let:
	\begin{equation}
	\hat{\chi}_{snap}(t) = ((1-\hat{M}_{B_1}(t)^2),...,(1-\hat{M}_{B_q}(t)^2)).
	\end{equation}
	Projecting the vector onto the full graph and then normalising by the magnitude allows us to define the \emph{survey-snapshot} influence strategy as:
	\begin{equation}
	\underline{h}_{snap}(t)=\frac{H}{H_{snap}}  G \hat{\chi}_{snap}(t),
	\end{equation}
	where $H_{snap}=\sum_{i=1}^N \big( G \hat{\chi}_{snap}\big)_i$.

\textbf{Null models.} In the $\underline{b}=0$ setting the obvious null model to contrast different influence strategies against is the uniform influence strategy. When $\underline{b}\ne0$ it is less clear what the appropriate null model is. We therefore explore two different approaches. The first, in which we target nodes (or blocks) with the lowest magnitude external fields will be referred to as the \emph{`hesitant targetting'} strategy. The second null model involves spending the budget to cancel out the negative external fields, we refer to this as the \emph{`negative canceller'} strategy. These strategies are defined in Table \ref{strat_summary} which also details the differing levels of information required by the different influence strategies.

\subsection{Influencing an Ising system on a large homophilous social network} \label{Pokec_influence_sect}

\begin{figure*}
	\centering
	
	\subfloat[]{\label{pokec_interactions}\includegraphics[width=0.31\textwidth]{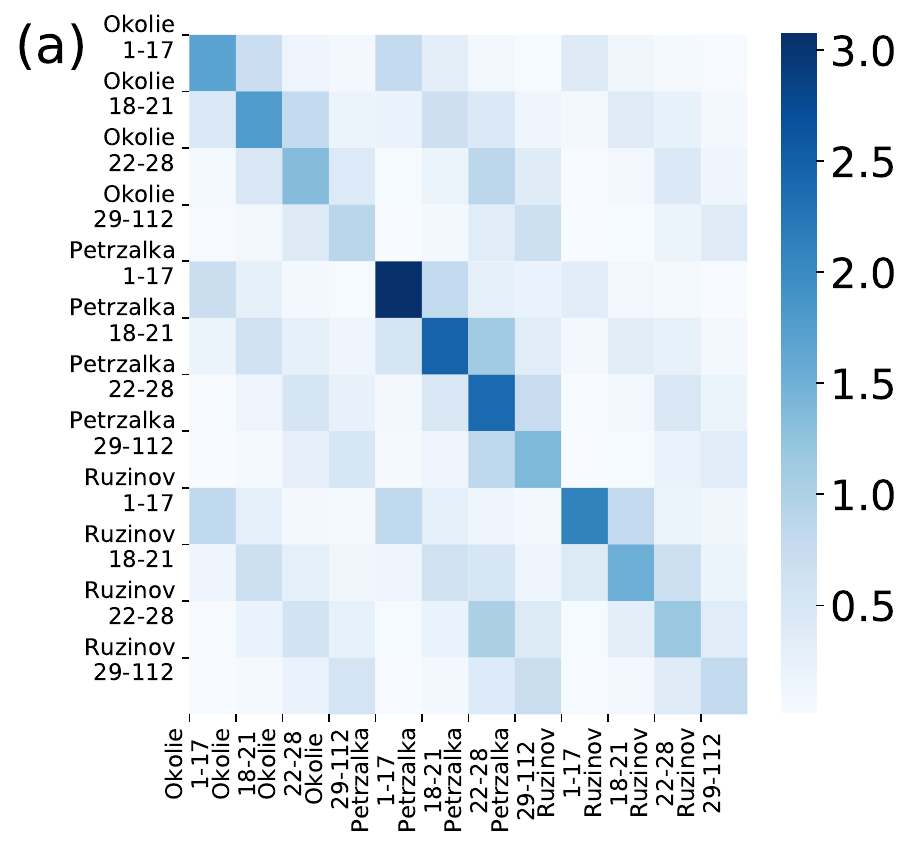}}
	\subfloat[]{\label{pokec_block_deg_dist}\includegraphics[width=0.31\textwidth]{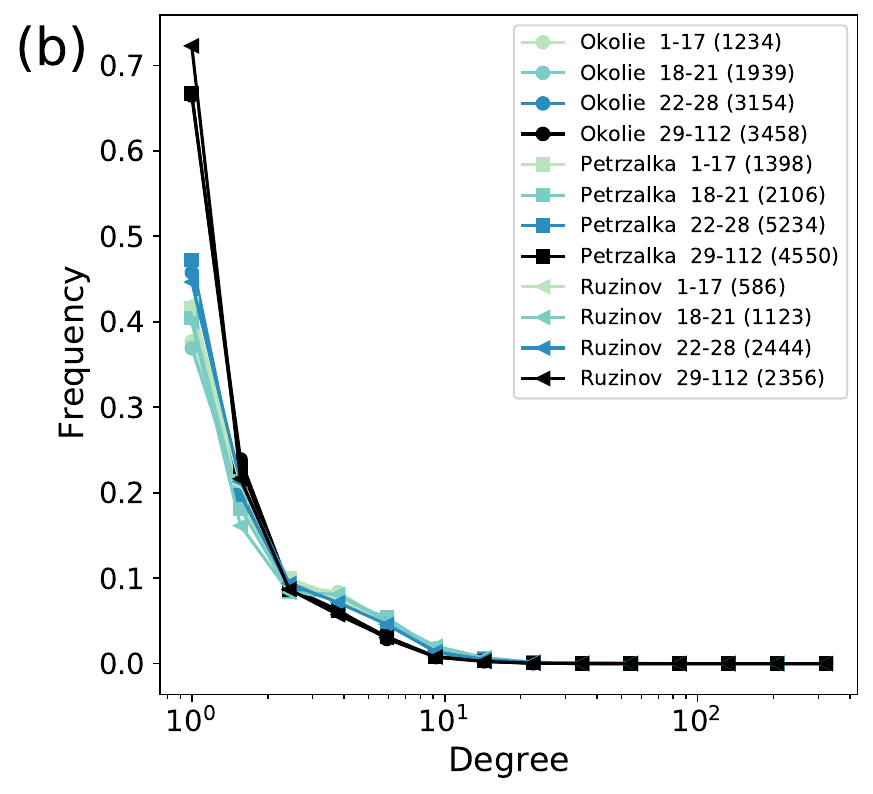}}
	\subfloat[]{\label{marks_up_as_beta}\includegraphics[width=0.31\textwidth]{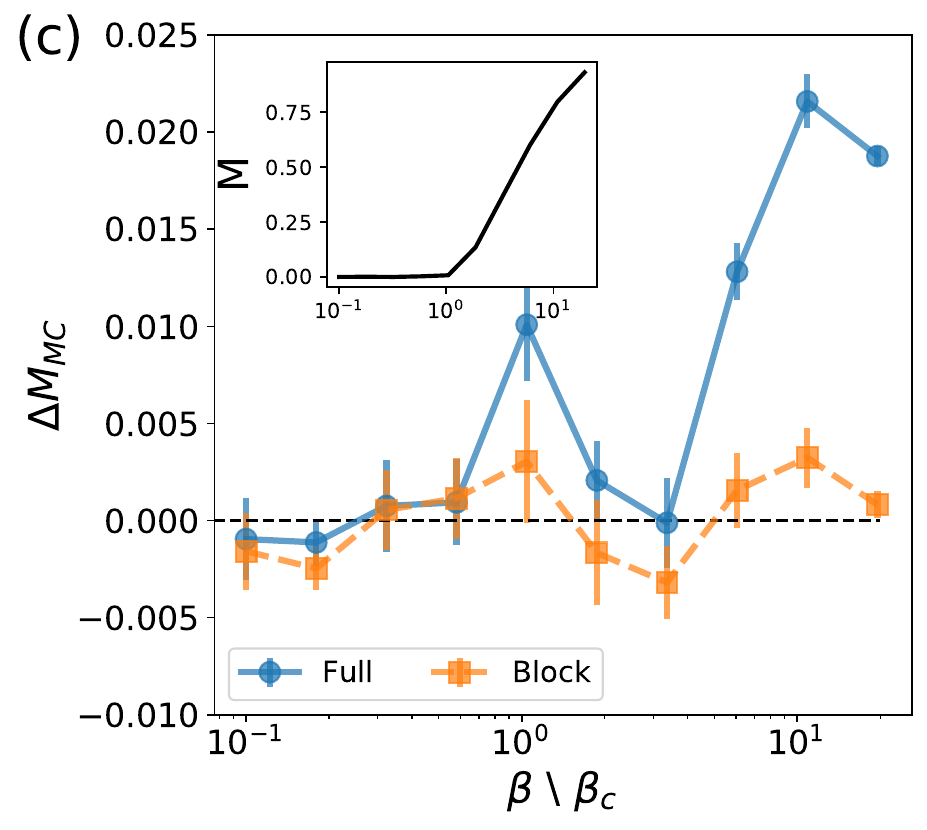}}
	
	\vspace*{-2.0em}
	\caption{\textbf{Without ambient fields, the heterogeneity in block degrees in the Pokec network is insufficient for block-level targetting to be effective for the chosen coarse-graining.} (a) Plot showing the elements of the the coupling matrix between different demographic blocks in the Pokec social network. The coupling matrix $\mathcal{K}$ was determined by counting the number of edges between the different blocks and inverting equations \ref{edges_from_con_prob} and \ref{sbm_coupling_matrix}. The large diagonal elements indicate strong homophily within blocks of the same age and region, while those on the off diagonals indicate age homophily between blocks. (b) Degree distributions at the level of blocks for the Pokec social network. The number of nodes in each block is reported in the legend. The proportions of nodes with each degree in the blocks are broadly comparable, with the exception of the tail of the distribution. (c) Plot showing how $\Delta M^{MC}$ varies as a function of $\beta$ for $H=2500$ for full graph (orange squares) and block-level (blue circles) influence strategies. Average magnetisation values were estimated by averaging over 15 Monte Carlo chains with a burn-in time of $6 \times 10^5$ steps and a run time of $10^4$ steps. The inset in (c) shows how the average magnetisation varies as a function of $\beta$ for the uniform control. Monte-Carlo simulations and fixed-point iteration (Algorithm \ref{mf_mag_alg}) are initialised at (1,1,..,1) with the intention of identifying the most positive metastable solution. Error bars indicate the standard error on the mean.}
	\label{pokec_no_field}
\end{figure*}

In this section we explore the performance of the different influence strategies defined above in a setting which engages with some aspects of real-world disorder. We apply the influence strategies to data obtained from an online social network. We consider a significantly larger network than previous ($N=29,582$) in order to demonstrate scalability of the influence strategies. The attributed network dataset was obtained from the online social network Pokec \cite{takac2012data} (see Supplementary Section \ref{pokec_data_description} for details). We extracted a sub-graph associated with nodes in the vicinity of Bratislava and coarse-grain the resulting network into twelve blocks based on the location and age of individuals (see Supplementary Section \ref{Pokec_block_extract}). We find strong homophily in both age and region (Figure \ref{pokec_interactions}), indicating that this form of coarse-graining provides an appropriate summary of the network structure.

\subsubsection{Impact of block-level degree heterogeneity ($\underline{b}=0$)} \label{pokec_deg_het}

In Section \ref{ising_scalable} we illustrated that, in the absence of ambient fields, the block-level influence strategy can perform well when there is heterogeneity in the average degrees. The Pokec network has a heterogeneous degree distribution \cite{takac2012data} and variability in the average degree at the level of blocks (Figure \ref{pokec_interactions}). However, the degree distributions at the level of blocks are broadly comparable (Figure \ref{pokec_block_deg_dist}). Consequently, we will expect poor performance from the block-level influence strategy at this level of coarse-graining. To test this we derive low budget influence strategies by computing $\nabla_{\underline{h}} M^{MF}$ and $\nabla_{\underline{h}_B} M^{MF}$ (see Equations \ref{linear_eq_for_sus} and \ref{block_mf_sus_equation} in the Supplement). Normalising by the sum of the elements and multiplying by $H$ in each case allows us to obtain a control with budget $H$. For our simulations we have selected $H=2500$. This value of $H$ is sufficient to observe changes in magnetisation of a few percent but sufficiently small to align with the assumption of low field budget. 

Figure \ref{marks_up_as_beta} shows the behaviour of $\Delta M^{MC}$ as a function of $\beta$ for the most positive metastable solution of the Pokec network. The block-level influence strategy performs poorly for a range of $\beta$ values indicating that this level of coarse-graining is not sufficient to effectively influence the system. On the other hand, it is possible to achieve a noticeable markup using the full graph influence strategy which acts as an upper bound on the performance of coarsened mean-field influence strategies. The largest markup occurs for large $\beta$ which agrees with previous observations that mean-field based influence strategies (such as $\underline{h}_{\mathrm{full}}$) are most effective away from the critical temperature \cite{lynn2016maximizing,lynn2017statistical}. These results hint that there might be finer levels of coarse-graining for which the block-level control does obtain an advantage.

Given the above, knowledge of the particular coarse-grained connectivity matrix $\mathcal{K}$ may not appear of much utility for influence Ising systems on this network in the high $\beta$ setting. However, in the case where $\underline{b}\ne0$, knowledge of $\mathcal{K}$ (combined with an appropriate initial condition) is required to estimate the susceptibility using Equation \ref{mf_equations_block_level}.

\begin{figure*}
	\centering
	\vspace*{-2.0em}
	\subfloat[]{\label{Pokec_block_mags_-1-0}\includegraphics[width=0.33\textwidth]{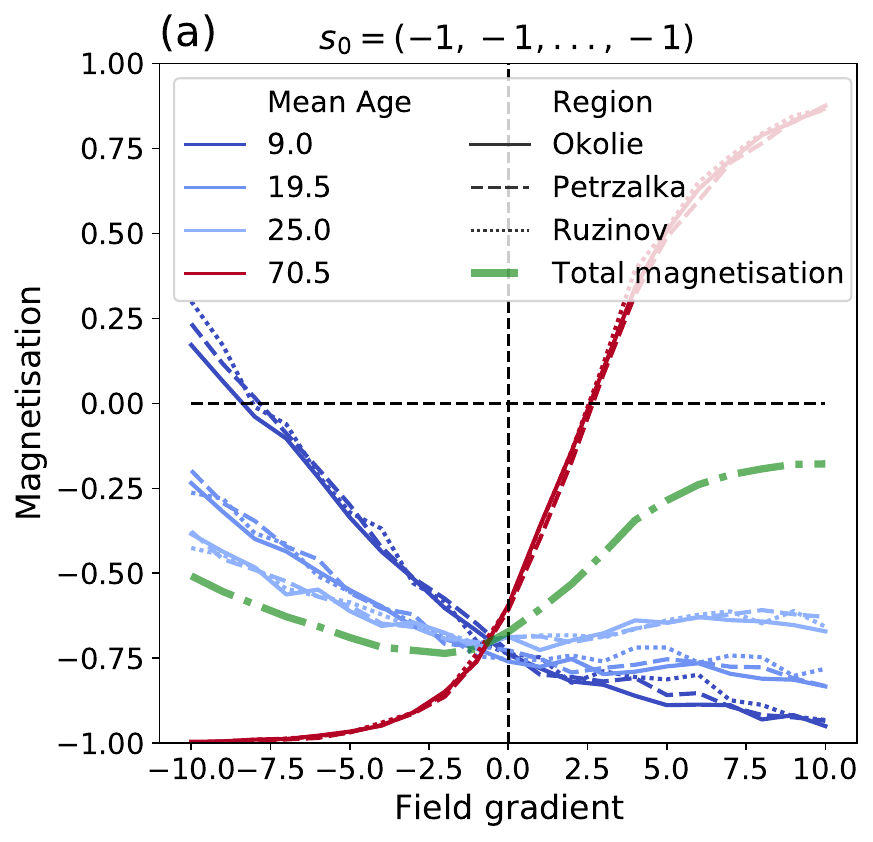}}
	\subfloat[]{\label{Pokec_block_mags_1-0}\includegraphics[width=0.33\textwidth]{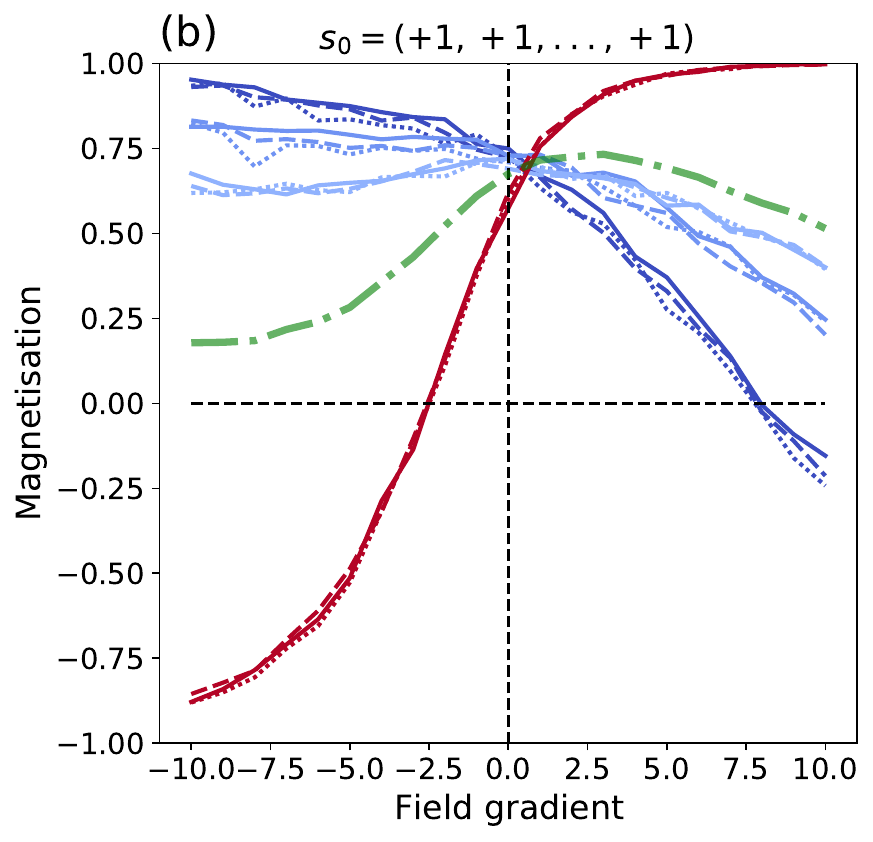}}
	
	\subfloat[]{\label{pokec_w_fields_magcontrols_pos}\includegraphics[width=0.33\textwidth]{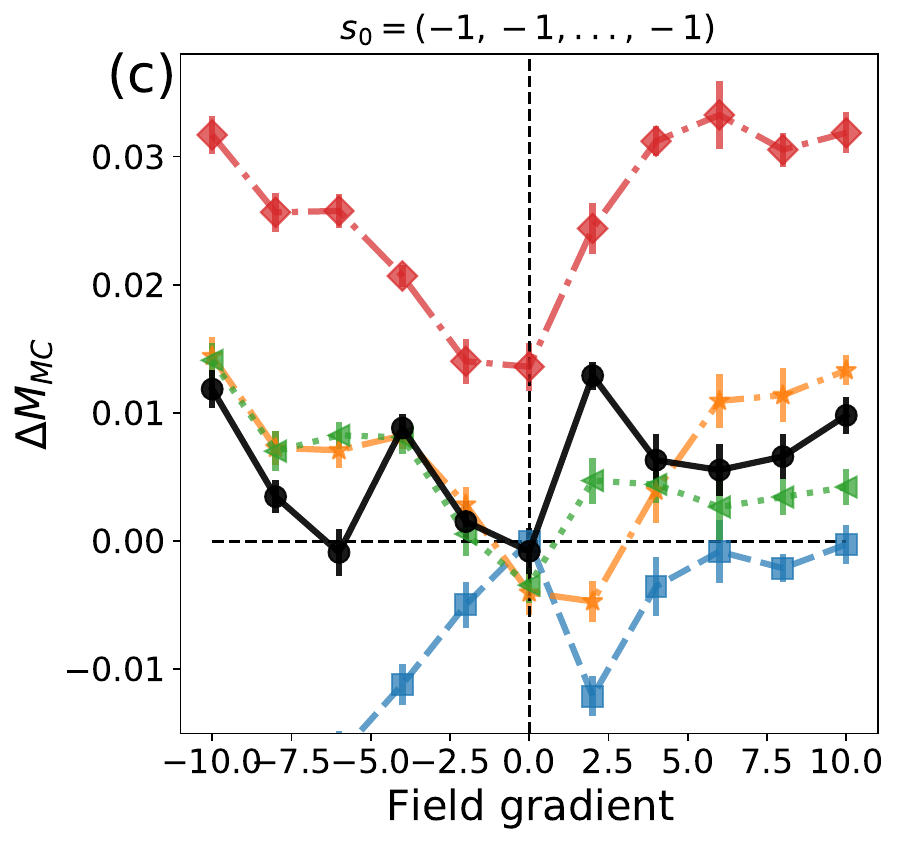}}
	\subfloat[]{\label{pokec_w_fields_magcontrols}\includegraphics[width=0.33\textwidth]{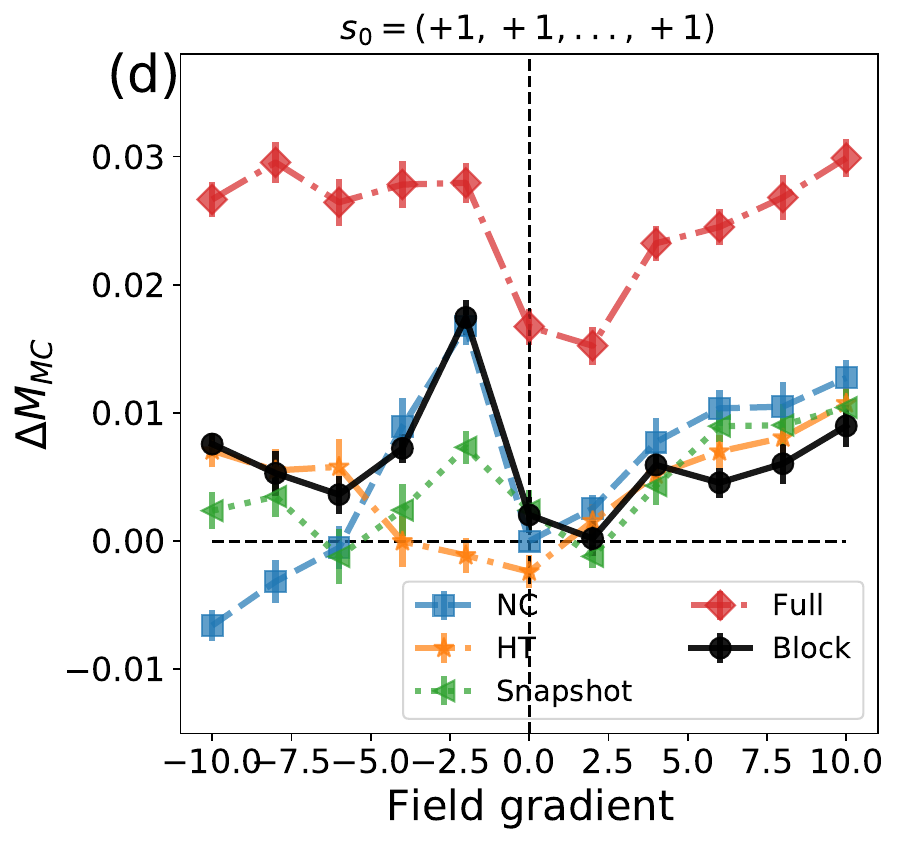}}
	
	\vspace*{-2.0em}
	\caption{\textbf{Knowledge of coarse-grained connectivity, ambient fields or point estimates of the magnetisation can be used to influence an Ising system on an online social network with ambient ambient fields.} Figures (a) and (b) show how the average magnetisation for different blocks varies as $g$ is increased for the positive and negative metastable solutions respectively. We also show the average magnetisation across all blocks (green dash-dotted line). Plots (c) and (d) show $\Delta M_{MC}$ for different influence strategies defined in Table \ref{strat_summary} as a function of the linear field gradient $g$ of the corresponding metastable solutions. Shown for full graph, block-level, hesitant targetting, negative canceller and survey-snapshot influence strategies for $H=2500$, $\beta = 8 \beta_c$, with $\beta_c$ being the inverse of the spectral radius of the full graph. In (a) and (b) we estimate the magnetisation by averaging over a single Monte Carlo chain with a burn-in time of $6 \times 10^5$ steps and a run time of $10^4$ steps. For (c) and (d) we take the average over 15 Monte Carlo chains with the same burn-in and run times. Error bars indicate the standard error on the mean.}
	\label{pokec_w_fields_magcontrols_both}
\end{figure*}

\subsubsection{Impact of adding external fields}

The Pokec social network dataset does not encode any ambient fields associated with the nodes. Consequently, we will impose an artificial ambient field to replicate what we might expect in a real system. Vaccine sentiment is known to covary with age in some countries \cite{de_figueiredo2016forecasted}. Motivated by this, we will consider an ambient field which varies linearly with age (see Supplementary Section \ref{Pokec_external_sim}) with gradient $g$. Using Monte Carlo simulations and the mean-field approximation we found that this system has two metastable solutions for a range of $\beta$ and $g$ values (See Supplementary Figures \ref{pokec_phase_diagram} and \ref{Pokec_robustness}). The block-level mean-field approximation is able to identify these metastable solutions when supplied with the same initial conditions, however, under certain conditions we find that it possesses additional spurious metastable states (see Supplementary Figure \ref{pokec_phase_diagram}). The positive and negative metastable solutions were obtained by initialising the fixed-point iteration and MC simulations at $(1,1,...,1)$ and $(-1,-1,...,-1)$ respectively. We validated that the system arrives in, and then stays in, these metastable solutions for an appreciable number of timesteps given the initial conditions (See Supplementary Figure \ref{Pokec_robustness}).

In this section we set $\beta=8 \beta_c$. This choice of $\beta$ corresponds to a `moderately' low temperature in which spins are broadly aligned with their blocks but still allows for the possibility for spins to flip due to being influenced or thermal fluctuations (See Supplementary Figure \ref{pokec_phase_diagram}). It also corresponds to the regime in sub-figure \ref{marks_up_as_beta} where it possible to obtain a noticeable increase in magnetisation using knowledge of the full graph.


The full and block-level mean-field based influence strategies are derived as described in Section \ref{pokec_deg_het} above. For the linear ambient field considered, the 22-28 age group has the lowest magnitude ambient field. We select this set of nodes to be $\mathcal{H}(\psi)$ when deriving the hesitant targetting influence strategy. We generate a single instance of the survey-snapshot control for each value of $g$. 

Figures \ref{Pokec_block_mags_1-0} and \ref{Pokec_block_mags_-1-0} show how the average magnetisation of the 12 blocks varies as the field gradient $g$ is increased. For $\beta = 8 \beta_c$ the magnitude of the average magnetisation is comparable for both metastable solutions when $g=0$ ($|M| \approx 0.7$). Increasing the magnitude of $g$ leads to greater polarisation. Inspecting Figures \ref{Pokec_block_mags_1-0} and \ref{Pokec_block_mags_-1-0} allows us to identify which blocks are least polarised (and therefore most susceptible to influence) for a given external field. Field gradients larger in magnitude than +/- 5 would be pertinent in a highly (possibly unrealistically) polarized society.

Figures \ref{pokec_w_fields_magcontrols} and \ref{pokec_w_fields_magcontrols_pos} show how the value of $\Delta M^{MC}$ for each strategy varies as a function of $g$ for the negative and positive metastable solutions respectively. In all cases, the control which uses knowledge of the full graph performs best. The order of the other controls varies for different values of $g$. The block-level IIM and survey-snapshot both perform consistently well across the parameter range considered. However, the hesitant targetting and negative canceller strategies also perform well in cases where the targetted blocks coincide with those with the lowest polarisation. We observes peaks in the performance of the block-level control when the magnetisation of the blocks with average age of 70.5 is approximately zero (this occurs for $g\approx2.5$ in Figure \ref{Pokec_block_mags_-1-0} and at $g\approx-2.5$ in Figure \ref{Pokec_block_mags_1-0}). This indicates that this strategy successfully identifies the least polarised group in the system.

Without external fields, the uniformity of the three-block system considered in Figure \ref{three_block_phase_diagram} would prevent us from being able to effectively influence the system with coarse-grained strategies. Figure \ref{pokec_no_field} provides evidence that the same is true for the chosen partition of the the Pokec social network. However, the presence of external fields allows us to achieve values of $\Delta M^{MC}$ of $\approx$50\% of those obtained by the full graph strategy relative to uniform targetting of blocks. It is worth noting that, although $\Delta M^{MC}\approx0$ for the block-level strategy for $g=0$, the success of the full graph mean-field suggests that it may be possible to achieve higher values with a finer partition. 

\emph{Impact of temperature.} We have not explicitly computed the impact of temperature. However, we expect it to be comparable to that in Section \ref{three_block_sbm_section}. As noted in the introduction, we believe that setting of having ambient fields which are sufficiently strong, at the given temperature, to align blocks is the most relevant. As the temperature or noise is increased ($\beta$ decreased), for fixed $\underline{b}$, all of the blocks will tend toward disorder (See Supplementary Figure \ref{pokec_phase_diagram}) and eventually the optimal block-level strategy will be relatively uniform. However, if the dataset under consideration has variability in the block-level degrees, it will be advantageous to target the highly connected nodes or blocks in the low $\beta$ limit as in Refs. \cite{lynn2016maximizing} and \cite{lynn2017statistical}.

\section{Discussion}

In this paper we have considered how we can influence Ising systems on modular social networks via weak but sustained perturbations in the external field. We considered influence strategies in the strong ambient field limit, which we suggest is the sociologically relevant setting in which to consider Ising systems. For small field budgets we found even node-level targeting strategies offered only a moderate opportunity to outperform the uniform baseline strategy. Nonetheless, it is possible, and the block-level influence strategy is a fruitful approach. We found that non-uniform Ising influence strategies can be effective when nodes have heterogeneous degrees (Figures \ref{two_block_markups} and Supplementary  Figure \ref{kappa_impact}) or experience different ambient fields (Figures \ref{three_block_phase_diagram} and \ref{pokec_w_fields_magcontrols_both}). We found various coarse-grained strategies can achieve a macroscopic fraction (up to $50 \%$) of the gain achieved using the full graph relative to a uniform baseline. Our findings indicate the possibility of effective public information campaigns which use tools from network science whilst having incomplete network data. Our model relied on a relatively strong set of assumptions. We will now outline some key limitations of our approach and directions for future research.

\emph {Robustness of metastable solutions.} In order to evaluate the impact of our `moderate time horizon' influence strategy in Figures 1,2 and 4 we considered initial conditions which consistently arrive in the same metastable solution. This behaviour is not representative of all initial conditions. For example, in the Pokec social network, initialising with a random initial condition with $M\approx0$ leads the system to arrive in one of the two metastable solutions with an appreciable frequency (see Supplementary Figure \ref{Pokec_robustness}). This issue would become more pronounced in systems where the phase space contains numerous metastable solutions, such as the spin glasses considered in Ref. \cite{aspelmeier2006free-energy}. In this case, an `infinite time horizon control' based approach may be required to obtain a robust influence strategy.

\emph{Background field correlates with module membership.} In Sections \ref{three_block_sbm_section} and \ref{Pokec_influence_sect} we assumed that the ambient background field experienced by nodes is equivalent for nodes in the same module. This assumption reflects the coarseness of the information we would obtain by performing inference of connectivity and external fields at the level of demographic groups using techniques such as those presented in Refs. \cite{godoy2021inference} or \cite{burioni2015enhancing}. In practice, the ambient external fields experienced by nodes may not be fully determined by their block membership. 

In Section \ref{Pokec_influence_sect} we selected node groups based on attributes. A more informed partition of the graph could be obtained using a community detection algorithm. Typically, this would require network information, however, it has been shown that community detection can be performed based on other forms of information, such as time series associated with the nodes \cite{hoffmann2020community,schaub2019blind,peixoto2019network}. We might also include more detailed information about the node degrees into influence strategies using using degree corrected SBMs \cite{karrer2011stochastic} or variants of the mean-field approximation \cite{suchecki2009bistable-monostable}. 

\emph{Impact of ensemble variability.} The simulations presented in this study have been performed on particular instances of SBMs. In future it will be of interest to explore the robustness of coarse-grained Ising influence strategies to randomness in the graph structure and how this depends on the choice of $\mathcal{K}$ and $\underline{b}$. Intuitively, the results may be relatively robust for larger systems. In Ref. \cite{hindes2017large} they link optimal regime switching strategies for Ising systems on networks to the principle eigenvector of the adjacency matrix. This suggests the possibility that results from random matrix theory in the context of modular networks (e.g. \cite{peixoto2013eigenvalue,kadavankandy2015characterization}) might be used to theoretically explore the ensemble averaged results.

In this study we consider partial network information in the form of discrete node coordinates. In the continuous case, connection probability information has been used to forecast dynamics on spatial networks via diffusion \cite{fisher2017social,smith2018using} or PDE based approaches \cite{hoffmann2018partially-observed,lang2017random}. Given a node distribution and connectivity kernel, the predictability of dynamical properties in spatial networks is impacted by dimensionality and inhomogeneity in the node distribution \cite{garrod2018large}. We expect our ability to influence dynamics on spatial networks to depend on the same factors. Exploring continuous systems may allow us to obtain a deeper understanding of factors which make it possible to influence Ising systems on networks without full network information. Our work provides a proof of concept and methodology for influencing opinion dynamics with noise and external covariates. We conclude by emphasising that while it can be insightful to blend simple physical models with realistically motivated ambient fields and network structure, the real world is clearly a richer dynamical system.

\section*{Data accessibility} Data used for Figures 3, 4 and S3-S6 is available from the Stanford Network Analysis Project (\href{https://snap.stanford.edu/data/soc-Pokec.html}{https://snap.stanford.edu/data/soc-Pokec.html}). Code to reproduce the results is available at
\href{https://github.com/MGarrod1/unobserved_spin_influence}{https://github.com/MGarrod1/unobserved\_spin\_influence}

\section*{Author's contributions} MG. Numerical simulations, data analysis, literature review, theoretical design, analytic computations and
manuscript writing. NJ. Interpretation of results, theoretical design and manuscript writing.

\section*{Funding} This work has been supported by EPSRC grants EP/L016613/1 and EP/N014529/1.

\section*{Acknowledgements} The authors would like to thank Chris Lynn, Sarab Sethi, Antonia Godoy-Lorite, Till Hoffman, Sahil Loomba and two anonymous referees for useful advice and comments on the manuscript.


\bibliographystyle{vancouver}
\bibliography{bibliography}

\begin{thebibliography}{10}

\bibitem{heckathorn1999aids}
Heckathorn DD, Broadhead RS, Anthony DL, Weakliem DL.
\newblock {AIDS} and social networks: {HIV} prevention through network
  mobilization.
\newblock Sociol Focus. 1999;32(2):159--179.
\newblock Available from: \url{https://doi.org/10.1080/00380237.1999.10571133}.

\bibitem{rice2012mobilizing}
Rice E, Tulbert E, Cederbaum J, Barman~Adhikari A, Milburn NG.
\newblock Mobilizing homeless youth for {HIV} prevention: a social network
  analysis of the acceptability of a face-to-face and online social networking
  intervention.
\newblock Health Educ Res. 2012;27(2):226--236.
\newblock Available from: \url{https://doi.org/10.1093/her/cyr113}.

\bibitem{vissenberg2016impact}
Vissenberg C, Stronks K, Nijpels G, Uitewaal PJM, Middelkoop BJC, Kohinor MJE,
  et~al.
\newblock Impact of a social network-based intervention promoting diabetes
  self-management in socioeconomically deprived patients: a qualitative
  evaluation of the intervention strategies.
\newblock BMJ Open. 2016;6(4):e010254.
\newblock Available from: \url{http://dx.doi.org/10.1136/bmjopen-2015-010254}.

\bibitem{vissenberg2017impact}
Vissenberg C, Nierkens V, Van~Valkengoed I, Nijpels G, Uitewaal P, Middelkoop
  B, et~al.
\newblock The impact of a social network based intervention on self-management
  behaviours among patients with type 2 diabetes living in socioeconomically
  deprived neighbourhoods: a mixed methods approach.
\newblock Scand J Public Healt. 2017;45(6):569--583.
\newblock Available from: \url{https://doi.org/10.1177%2F1403494817701565}.

\bibitem{vissenberg2017development}
Vissenberg C, Nierkens V, Uitewaal PJM, Middelkoop BJC, Nijpels G, Stronks K.
\newblock Development of the social network-{Based} intervention “{Powerful}
  {Together} with {Diabetes}” {Using} intervention {Mapping}.
\newblock Front Public Health. 2017;5:334.
\newblock Available from: \url{https://doi.org/10.3389/fpubh.2017.00334}.

\bibitem{pechmann2017randomised}
Pechmann C, Delucchi K, Lakon CM, Prochaska JJ.
\newblock Randomised controlled trial evaluation of {Tweet2Quit}: a social
  network quit-smoking intervention.
\newblock Tob Control. 2017;26(2):188--194.
\newblock Available from:
  \url{http://dx.doi.org/10.1136/tobaccocontrol-2015-052768}.

\bibitem{tsoh2015social}
Tsoh JY, Burke NJ, Gildengorin G, Wong C, Le K, Nguyen A, et~al.
\newblock A social network family-focused intervention to promote smoking
  cessation in {Chinese} and {Vietnamese} {American} male smokers: a
  feasibility study.
\newblock Nicotine Tob Res. 2015;17(8):1029--1038.
\newblock Available from: \url{https://doi.org/10.1093/ntr/ntv088}.

\bibitem{williams2015network}
Williams HTP, McMurray JR, Kurz T, Lambert FH.
\newblock Network analysis reveals open forums and echo chambers in social
  media discussions of climate change.
\newblock Global Environ Change. 2015;32:126--138.
\newblock Available from:
  \url{https://doi.org/10.1016/j.gloenvcha.2015.03.006}.

\bibitem{kaiser2017alliance}
Kaiser J, Puschmann C.
\newblock Alliance of antagonism: {Counterpublics} and polarization in online
  climate change communication.
\newblock Communication and the Public. 2017;2(4):371--387.
\newblock Available from: \url{https://doi.org/10.1177%2F2057047317732350}.

\bibitem{kahan2012polarizing}
Kahan DM, Peters E, Wittlin M, Slovic P, Ouellette LL, Braman D, et~al.
\newblock The polarizing impact of science literacy and numeracy on perceived
  climate change risks.
\newblock Nat Clim Change. 2012;2(10):732.
\newblock Available from: \url{https://doi.org/10.1038/nclimate1547}.

\bibitem{schmidt2018polarization}
Schmidt AL, Zollo F, Scala A, Betsch C, Quattrociocchi W.
\newblock Polarization of the vaccination debate on {Facebook}.
\newblock Vaccine. 2018;36(25):3606--3612.
\newblock Available from: \url{https://doi.org/10.1016/j.vaccine.2018.05.040}.

\bibitem{kempe2003maximizing}
Kempe D, Kleinberg J, Tardos E.
\newblock Maximizing the spread of influence through a social network.
\newblock In: Proceedings of the ninth {ACM} {SIGKDD} international conference
  on {Knowledge} discovery and data mining. ACM; 2003. p. 137--146.
\newblock Available from: \url{https://doi.org/10.1145/956750.956769}.

\bibitem{chen2010scalable}
Chen W, Yuan Y, Zhang L.
\newblock Scalable influence maximization in social networks under the linear
  threshold model.
\newblock In: 2010 {IEEE} international conference on data mining. IEEE; 2010.
  p. 88--97.
\newblock Available from: \url{https://doi.org/10.1109/ICDM.2010.118}.

\bibitem{wang2012scalable}
Wang C, Chen W, Wang Y.
\newblock Scalable influence maximization for independent cascade model in
  large-scale social networks.
\newblock Data Min Knowl Disc. 2012;25(3):545--576.
\newblock Available from: \url{https://doi.org/10.1007/s10618-012-0262-1}.

\bibitem{lynn2016maximizing}
Lynn C, Lee DD.
\newblock Maximizing influence in an ising network: {A} mean-field optimal
  solution.
\newblock In: Adv. Neur. In.; 2016. p. 2495--2503.

\bibitem{lynn2017statistical}
Lynn CW, Lee DD.
\newblock Statistical mechanics of influence maximization with thermal noise.
\newblock Europhys Lett. 2017;117(6):66001.
\newblock Available from: \url{https://doi.org/10.1209/0295-5075/117/66001}.

\bibitem{hindes2017large}
Hindes J, Schwartz IB.
\newblock Large order fluctuations, switching, and control in complex networks.
\newblock Sci Rep-UK. 2017;7.
\newblock Available from: \url{https://doi.org/10.1038/s41598-017-08828-8}.

\bibitem{lynn2018maximizing}
Lynn CW, Lee DD.
\newblock Maximizing {Activity} in {Ising} {Networks} via the {TAP}
  {Approximation}.
\newblock arXiv preprint arXiv:180300110. 2018;Available from:
  \url{https://arxiv.org/abs/1803.00110}.

\bibitem{mackay2003information}
MacKay DJC.
\newblock In: Information theory, inference and learning algorithms. Cambridge
  University Press, Cambridge, UK; 2003. p. 400,422--428.

\bibitem{newman1999monte}
Newman MEJ, Barkema G.
\newblock In: Monte Carlo methods in statistical physics. Oxford University
  Press: New York, USA; 1999. p. 13--16,46--53.

\bibitem{castellano2009statistical}
Castellano C, Fortunato S, Loreto V.
\newblock Statistical physics of social dynamics.
\newblock Rev Mod Phys. 2009;81(2):591.
\newblock Available from: \url{https://doi.org/10.1103/RevModPhys.81.591}.

\bibitem{acemoglu2011opinion}
Acemoglu D, Ozdaglar A.
\newblock Opinion dynamics and learning in social networks.
\newblock Dyn Games Appl. 2011;1(1):3--49.
\newblock Available from: \url{https://doi.org/10.1007/s13235-010-0004-1}.

\bibitem{weber2016cellular}
Weber M, Buceta J.
\newblock The cellular Ising model: a framework for phase transitions in
  multicellular environments.
\newblock Journal of The Royal Society Interface. 2016;13(119):20151092.
\newblock Available from: \url{https://doi.org/10.1098/rsif.2015.1092}.

\bibitem{lynn2019physics}
Lynn CW, Bassett DS.
\newblock The physics of brain network structure, function and control.
\newblock Nat Rev Phys. 2019;1(5):318.
\newblock Available from: \url{https://doi.org/10.1038/s42254-019-0040-8}.

\bibitem{nareddy2020dynamical}
Nareddy VR, Machta J, Abbott KC, Esmaeili S, Hastings A.
\newblock Dynamical Ising model of spatially coupled ecological oscillators.
\newblock J R Soc Interface. 2020;17(171):20200571.
\newblock Available from: \url{https://doi.org/10.1098/rsif.2020.0571}.

\bibitem{kindermann1980markov}
Kindermann R, Snell JL.
\newblock Markov random fields and their applications.
\newblock Amer Math Soc. 1980;.

\bibitem{possolo1986estimation}
Possolo A.
\newblock Estimation of binary {Markov} random fields.
\newblock Technical Repot. 1986;(77).

\bibitem{cressie1992statistics}
Cressie N.
\newblock Statistics for spatial data.
\newblock Terra Nova. 1992;4(5):613--617.
\newblock Available from:
  \url{https://doi.org/10.1111/j.1365-3121.1992.tb00605.x}.

\bibitem{besag1974spatial}
Besag J.
\newblock Spatial interaction and the statistical analysis of lattice systems.
\newblock J Roy Stat Soc B Met. 1974;p. 192--236.
\newblock Available from:
  \url{https://doi.org/10.1111/j.2517-6161.1974.tb00999.x}.

\bibitem{tanaka1998mean-field}
Tanaka T.
\newblock Mean-field theory of {Boltzmann} machine learning.
\newblock Phys Rev E. 1998;58(2):2302.
\newblock Available from: \url{https://doi.org/10.1103/PhysRevE.58.2302}.

\bibitem{godoy2021inference}
Godoy-Lorite A, Jones NS.
\newblock Inference and influence of network structure using snapshot social
  behavior without network data.
\newblock Sci Adv. 2021;7(23):eabb8762.
\newblock Available from: \url{https://doi.org/10.1126/sciadv.abb8762}.

\bibitem{burioni2015enhancing}
Burioni R, Contucci P, Fedele M, Vernia C, Vezzani A.
\newblock Enhancing participation to health screening campaigns by group
  interactions.
\newblock Sci Rep. 2015;5:9904.
\newblock Available from: \url{https://doi.org/10.1038/srep09904}.

\bibitem{day2019glassy}
Day AGR, Bukov M, Weinberg P, Mehta P, Sels D.
\newblock Glassy phase of optimal quantum control.
\newblock Phys Rev Lett. 2019;122(2):020601.
\newblock Available from: \url{https://doi.org/10.1103/PhysRevLett.122.020601}.

\bibitem{rotskoff2017geometric}
Rotskoff GM, Crooks GE, Vanden-Eijnden E.
\newblock Geometric approach to optimal nonequilibrium control: {Minimizing}
  dissipation in nanomagnetic spin systems.
\newblock Phys Rev E. 2017;95(1):012148.
\newblock Available from: \url{https://doi.org/10.1103/PhysRevE.95.012148}.

\bibitem{hoffmann2020inference}
Hoffmann T, Jones NS.
\newblock Inference of a universal social scale and segregation measures using
  social connectivity kernels.
\newblock J R Soc Interface. 2020;17:20200638.
\newblock Available from: \url{https://doi.org/10.1098/rsif.2020.0638}.

\bibitem{mcpherson2019network}
McPherson M, Smith JA.
\newblock Network {Effects} in {Blau} {Space}: {Imputing} {Social} {Context}
  from {Survey} {Data}.
\newblock Socius. 2019;5:2378023119868591.
\newblock Available from: \url{https://doi.org/10.1177%2F2378023119868591}.

\bibitem{butts2011spatial}
Butts CT, Acton RM.
\newblock Spatial modeling of social networks.
\newblock The Sage Handbook of GIS and Society Research Thousand Oaks, CA: SAGE
  Publications. 2011;p. 222--250.

\bibitem{blau1977macrosociological}
Blau PM.
\newblock A macrosociological theory of social structure.
\newblock Am J of Sociol. 1977;83(1):26--54.
\newblock Available from: \url{https://doi.org/10.1086/226505}.

\bibitem{hoffmann2018partially-observed}
Hoffmann T.
\newblock Partially-observed networks: {Inference} and dynamics [{PhD}
  {Thesis}]; 2018.
\newblock Available from: \url{https://doi.org/10.25560/71325}.

\bibitem{smith2019continued}
Smith JA.
\newblock The {Continued} {Relevance} of {Ego} {Network} {Data}. 2019;Available
  from: \url{https://doi.org/10.31235/osf.io/phfvq}.

\bibitem{smith2015global}
Smith JA.
\newblock Global network inference from ego network samples: testing a
  simulation approach.
\newblock J Math Sociol. 2015;39(2):125--162.
\newblock Available from: \url{https://doi.org/10.1080/0022250X.2014.994621}.

\bibitem{mccormick2015latent}
McCormick TH, Zheng T.
\newblock Latent surface models for networks using {Aggregated} {Relational}
  {Data}.
\newblock J Am Stat Assoc. 2015;110(512):1684--1695.
\newblock Available from: \url{https://doi.org/10.1080/01621459.2014.991395}.

\bibitem{breza2017using}
Breza E, Chandrasekhar AG, McCormick TH, Pan M.
\newblock Using {Aggregated} {Relational} {Data} to feasibly identify network
  structure without network data.
\newblock Nat. Bur. Econ. Res. Working Paper Series; 2017.
\newblock Available from: \url{https://doi.org/10.3386/w23491}.

\bibitem{breza2019consistently}
Breza E, Chandrasekhar AG, McCormick TH, Pan M.
\newblock Consistently estimating graph statistics using {Aggregated}
  {Relational} {Data}.
\newblock arXiv preprint arXiv:190809881. 2019;Available from:
  \url{https://arxiv.org/abs/1908.09881}.

\bibitem{gallo2009parameter}
Gallo I, Barra A, Contucci P.
\newblock Parameter evaluation of a simple mean-field model of social
  interaction.
\newblock Math Mod Meth Appl S. 2009;19(supp01):1427--1439.
\newblock Available from: \url{https://doi.org/10.1142/S0218202509003863}.

\bibitem{gallo2009equilibrium}
Gallo I.
\newblock An equilibrium approach to modelling social interaction.
\newblock arXiv preprint arXiv:09072561. 2009;Available from:
  \url{https://arxiv.org/abs/0907.2561}.

\bibitem{opoku2019parameter}
Opoku AA, Osabutey G, Kwofie C.
\newblock Parameter Evaluation for a Statistical Mechanical Model for Binary
  Choice with Social Interaction.
\newblock Journal of Probability and Statistics. 2019;2019.
\newblock Available from: \url{https://doi.org/10.1155/2019/3435626}.

\bibitem{fedele2013inverse}
Fedele M, Vernia C, Contucci P.
\newblock Inverse problem robustness for multi-species mean-field spin models.
\newblock J Phys A-Math Theor. 2013;46(6):065001.
\newblock Available from: \url{https://doi.org/10.1088/1751-8113/46/6/065001}.

\bibitem{fedele2017inverse}
Fedele M, Vernia C.
\newblock Inverse problem for multispecies ferromagneticlike mean-field models
  in phase space with many states.
\newblock Phys Rev E. 2017;96(4):042135.
\newblock Available from: \url{https://doi.org/10.1103/PhysRevE.96.042135}.

\bibitem{peixoto2019network}
Peixoto TP.
\newblock Network reconstruction and community detection from dynamics.
\newblock Phys Rev Lett. 2019;123(12):128301.
\newblock Available from: \url{https://doi.org/10.1103/PhysRevLett.123.128301}.

\bibitem{schaub2019blind}
Schaub MT, Segarra S, Tsitsiklis JN.
\newblock Blind identification of stochastic block models from dynamical
  observations.
\newblock arXiv preprint arXiv:190509107. 2019;Available from:
  \url{https://arxiv.org/pdf/1905.09107}.

\bibitem{mcpherson2001birds}
McPherson M, Smith-Lovin L, Cook JM.
\newblock Birds of a feather: {Homophily} in social networks.
\newblock Annu Rev Sociol. 2001;p. 415--444.
\newblock Available from: \url{https://doi.org/10.1146/annurev.soc.27.1.415}.

\bibitem{smith2014social}
Smith JA, McPherson M, Smith-Lovin L.
\newblock Social distance in the {United} {States}: {Sex}, race, religion, age,
  and education homophily among confidants, 1985 to 2004.
\newblock Am Sociol Rev. 2014;79(3):432--456.
\newblock Available from: \url{https://doi.org/10.1177%2F0003122414531776}.

\bibitem{karrer2011stochastic}
Karrer B, Newman MEJ.
\newblock Stochastic blockmodels and community structure in networks.
\newblock Phys Rev E. 2011;83(1):016107.
\newblock Available from: \url{https://doi.org/10.1103/PhysRevE.83.016107}.

\bibitem{holland1983stochastic}
Holland PW, Laskey KB, Leinhardt S.
\newblock Stochastic blockmodels: {First} steps.
\newblock Soc Networks. 1983;5(2):109--137.
\newblock Available from: \url{https://doi.org/10.1016/0378-8733(83)90021-7}.

\bibitem{faust1992blockmodels}
Faust K, Wasserman S.
\newblock Blockmodels: {Interpretation} and evaluation.
\newblock Soc Networks. 1992;14(1-2):5--61.
\newblock Available from: \url{https://doi.org/10.1016/0378-8733(92)90013-W}.

\bibitem{aspelmeier2006free-energy}
Aspelmeier T, Blythe RA, Bray AJ, Moore MA.
\newblock Free-energy landscapes, dynamics, and the edge of chaos in mean-field
  models of spin glasses.
\newblock Phys Rev B. 2006;74(18):184411.
\newblock Available from: \url{https://doi.org/10.1103/PhysRevB.74.184411}.

\bibitem{suchecki2009bistable-monostable}
Suchecki K, Ho{\l}yst JA.
\newblock Bistable-monostable transition in the {Ising} model on two connected
  complex networks.
\newblock Phys Rev E. 2009;80(3):031110.
\newblock Available from: \url{https://doi.org/10.1103/PhysRevE.80.031110}.

\bibitem{lambiotte2007coexistence}
Lambiotte R, Ausloos M.
\newblock Coexistence of opposite opinions in a network with communities.
\newblock J Stat Mech: Theory Exp. 2007;2007(08):P08026.
\newblock Available from:
  \url{https://doi.org/10.1088/1742-5468/2007/08/P08026}.

\bibitem{lambiotte2007majority}
Lambiotte R, Ausloos M, Ho{\l}yst JA.
\newblock Majority model on a network with communities.
\newblock Phys Rev E. 2007;75(3):030101.
\newblock Available from: \url{https://doi.org/10.1103/PhysRevE.75.030101}.

\bibitem{opoku2018multipopulation}
Opoku AA, Osabutey G.
\newblock Multipopulation spin models: a view from large deviations theoretic
  window.
\newblock J Math. 2018;2018.
\newblock Available from: \url{https://doi.org/10.1155/2018/9417547}.

\bibitem{suchecki2006ising}
Suchecki K, Ho{\l}yst JA.
\newblock Ising model on two connected {Barab{\'a}si}-{Albert} networks.
\newblock Phys Rev E. 2006;74(1):011122.
\newblock Available from: \url{https://doi.org/10.1103/PhysRevE.74.011122}.

\bibitem{lowe2019multi-group}
L{\"o}we M, Schubert K, Vermet F.
\newblock Multi-group binary choice with social interaction and a random
  communication structure: A random graph approach.
\newblock Physica A. 2020;556:124735.
\newblock Available from: \url{https://doi.org/10.1016/j.physa.2020.124735}.

\bibitem{opper2001advanced}
Opper M, Saad D.
\newblock Advanced mean field methods: {Theory} and practice.
\newblock MIT Press, Cambridge, MA; 2001.

\bibitem{yedidia2001idiosyncratic}
Yedidia J.
\newblock An idiosyncratic journey beyond mean field theory.
\newblock Advanced mean field methods: Theory and practice. 2001;p. 21--36.

\bibitem{butler2015diagnosing}
Butler R, MacDonald NE.
\newblock Diagnosing the determinants of vaccine hesitancy in specific
  subgroups: The Guide to Tailoring Immunization Programmes (TIP).
\newblock Vaccine. 2015;33(34):4176--4179.
\newblock Available from: \url{https://doi.org/10.1016/j.vaccine.2015.04.038}.

\bibitem{takac2012data}
Takac L, Zabovsky M.
\newblock Data analysis in public social networks.
\newblock In: International {Scientific} {Conference} and {International}
  {Workshop} {Present} {Day} {Trends} of {Innovations}. vol.~1; 2012. Available
  from: \url{https://www.mmds.org/data/soc-pokec.pdf}.

\bibitem{de_figueiredo2016forecasted}
de~Figueiredo A, Johnston IG, Smith DMD, Agarwal S, Larson HJ, Jones NS.
\newblock Forecasted trends in vaccination coverage and correlations with
  socioeconomic factors: a global time-series analysis over 30 years.
\newblock Lancet Glob Health. 2016;4(10):e726--e735.
\newblock Available from: \url{https://doi.org/10.1016/S2214-109X(16)30167-X}.

\bibitem{hoffmann2020community}
Hoffmann T, Peel L, Lambiotte R, Jones NS.
\newblock Community detection in networks without observing edges.
\newblock Sci Adv. 2020;6(4):eaav1478.
\newblock Available from: \url{https://doi.org/10.1126/sciadv.aav1478}.

\bibitem{peixoto2013eigenvalue}
Peixoto TP.
\newblock Eigenvalue spectra of modular networks.
\newblock Phys Rev Lett. 2013;111(9):098701.
\newblock Available from: \url{https://doi.org/10.1103/PhysRevLett.111.098701}.

\bibitem{kadavankandy2015characterization}
Kadavankandy A, Cottatellucci L, Avrachenkov K.
\newblock Characterization of random matrix eigenvectors for stochastic block
  model.
\newblock In: Proc. 49th Asilomar Conf. Signals, Syst. Comput. IEEE; 2015. p.
  861--865.
\newblock Available from: \url{https://doi.org/10.1109/ACSSC.2015.7421258}.

\bibitem{fisher2017social}
Fisher JC.
\newblock Social {Space} {Diffusion}: {Applications} of a {Latent} {Space}
  {Model} to {Diffusion} with {Uncertain} {Ties}.
\newblock Sociol Methodol. 2017;p. 0081175018820075.
\newblock Available from: \url{https://doi.org/10.1177%2F0081175018820075}.

\bibitem{smith2018using}
Smith JA, Burow J.
\newblock Using ego network data to inform agent-based models of diffusion.
\newblock Sociol Method Res. 2018;p. 0049124118769100.
\newblock Available from: \url{https://doi.org/10.1177%2F0049124118769100}.

\bibitem{lang2017random}
Lang J, De~Sterck H, Kaiser JL, Miller JC.
\newblock Random {Spatial} {Networks}: {Small} {Worlds} without {Clustering},
  {Traveling} {Waves}, and {Hop}-and-{Spread} {Disease} {Dynamics}.
\newblock arXiv preprint arXiv:170201252. 2017;Available from:
  \url{https://arxiv.org/abs/1702.01252}.

\bibitem{garrod2018large}
Garrod M, Jones NS.
\newblock Large algebraic connectivity fluctuations in spatial network
  ensembles imply a predictive advantage from node location information.
\newblock Phys Rev E. 2018;98(5).
\newblock Available from: \url{https://doi.org/10.1103/PhysRevE.98.052316}.

\bibitem{dahmen2016correlated}
Dahmen D, Bos H, Helias M.
\newblock Correlated fluctuations in strongly coupled binary networks beyond
  equilibrium.
\newblock Phys Rev X. 2016;6(3):031024.
\newblock Available from: \url{https://doi.org/10.1103/PhysRevX.6.031024}.

\bibitem{zou2011improved}
Zou W, Filatov M, Cremer D.
\newblock An improved algorithm for the normalized elimination of the
  small-component method.
\newblock Theor Chem Acc. 2011;130(4-6):633--644.
\newblock Available from: \url{https://doi.org/10.1007/s00214-011-1007-8}.

\bibitem{berinde2007iterative}
Berinde V.
\newblock Iterative Approximation of Fixed Points. vol. 1912 of Lecture Notes
  in Mathematics.
\newblock Berlin, Heidelberg: Springer, Berlin, Heidelberg; 2007.

\bibitem{bolthausen2014iterative}
Bolthausen E.
\newblock An iterative construction of solutions of the TAP equations for the
  Sherrington–-Kirkpatrick model.
\newblock Commun Math Phys. 2014;325(1):333--366.
\newblock Available from: \url{https://doi.org/10.1007/s00220-013-1862-3}.

\bibitem{nocedal2006numerical}
Nocedal J, Wright S.
\newblock In: Numerical Optimization. Springer Series in Operations Research
  and Financial Engineering. New York, NY: Springer New York; 2006. p. 30--41.

\bibitem{blondel2014large-scale}
Blondel M, Fujino A, Ueda N.
\newblock Large-scale multiclass support vector machine training via Euclidean
  projection onto the simplex.
\newblock In: 2014 22nd {International} {Conference} on {Pattern}
  {Recognition}. IEEE; 2014. p. 1289--1294.
\newblock Available from: \url{https://doi.org/10.1109/ICPR.2014.231}.

\bibitem{brooks2009contraction}
Brooks RM, Schmitt K.
\newblock The Contraction Mapping Principle and some Applications.
\newblock Electron J Differ Equ. 2009;.

\end{thebibliography}

\onecolumngrid

\pagebreak
\begin{center} 
	\vspace{20mm}
	\textbf{\large Influencing dynamics on social networks without knowledge of network microstructure}\\ \vspace{5mm} \large Supplementary material \\ \vspace{5mm}
	\large Matthew Garrod, Nick S. Jones \\ \vspace{5mm}
	\large \today
\end{center}


\beginsupplement

\section{Table of Symbols} \label{symbols}

The table \ref{symbols_table} below shows the definitions of the symbols used in the main text.

\begin{table}[]
	\resizebox{0.95\textwidth}{!}{%
	\begin{tabular}{|l|l|l|}
		\hline
		Symbol                                             & Meaning                                                                                                                & Domain                      \\ \hline
		$\underline{s}$                                    & Spin configuration for an Ising system on $N$ nodes.                                                                   & $\{-1,1\}^{N}$              \\ \hline
		$\mathcal{E}$                                      & Energy of the spin system.                                                                                             & $\mathbb{R}$                \\ \hline
		$A$                                                & Adjacency matrix for network.                                                                                          & $\{0,1\}^{N \times N}$      \\ \hline
		$\underline{g}$                                    & Total external field vector. Element $g_i$ is the total external field acting on node $i$.                             & $\mathbb{R}^N$              \\ \hline
		$\underline{h}$                                    & Controllable part of the total external field.                                                                         & $\mathbb{R}^N$              \\ \hline
		$\underline{b}$                                    & Ambient (uncontrollable) part of the total external field.                                                             & $\mathbb{R}^N$              \\ \hline
		$\mathbb{P}(x)$                                    & Probability of $x$.                                                                                                    & $[0,1]$                     \\ \hline
		$\beta$                                            & Inverse temperature of an Ising system.                                                                                & $\mathbb{R}^+$              \\ \hline
		$N$                                                & Number of nodes in a network.                                                                                          & $\mathbb{Z^+}$              \\ \hline
		$M$                                                & Magnetisation: $\frac{1}{N} \sum_{i=1}^N \langle s_i \rangle$.                                                         & $[-1,1]$                    \\ \hline
		$q$                                                & Number of blocks in SBM.                                                                                               & $\mathbb{Z^+}$              \\ \hline
		$\underline{t}$                                    & Block membership vector for SBM. $t_i=k$ if node $i$ is in block $k$.                                                  & $\{1,...,q\}^{N}$           \\ \hline
		$\Omega$                                           & $q \times q$ affinity or connection probability matrix for SBM.                                                        & $[0,1]^{q \times q}$        \\ \hline
		$\mathcal{K}$                                      & SBM coupling matrix. $\mathcal{K}_{xy}$ is the expected number of links between two nodes in blocks $x$ and $y$.       & $\mathbb{R}^{q \times q}$   \\ \hline
		$\mathcal{N}$                                      & Diagonal matrix of SBM block sizes.                                                                                    & $\mathbb{Z^+}^{q \times q}$ \\ \hline
		$E$                                                & $E_{xy}$ is the expected number of edges between blocks $x$ and $y$.                                                   & $\mathbb{R}^{q \times q}$   \\ \hline
		$\beta_c$                                          & Critical inverse temperature of an Ising system ($A$, $\underline{b}$) computed using the mean-field approximation.    & $\mathbb{R}^+$              \\ \hline
		$H$                                                & Field budget for Ising influence strategies.                                                                           & $\mathbb{R}^+$              \\ \hline
		$\underline{h}^*$                                  & Ising influence strategy which maximises the magnetisation.                                                            & $\mathbb{R}^N$              \\ \hline
		$\underline{m}$                                    & Mean-field magnetisation vector.                                                                                       & $[-1,1]^N$                  \\ \hline
		$M^{MF}$                                           & Average mean-field mangetisation $M^{MF} = \frac{1}{N} \sum_{i=1}^{N} m_i$.                                            & $[-1,1]$                    \\ \hline
		$\nabla_{\underline{h}} M^{MF}|_{\underline{h}=0}$ & Mean-field magnetisation gradient (susceptibility vector). See Equation \ref{Gradient_mag_def}.                        & $\mathbb{R}^N$              \\ \hline
		$\underline{m}_{B}$                                & Block-level average mean-field magnetisation vector.                                                                   & $[-1,1]$                    \\ \hline
		$\underline{\tilde{b}}_{B}$                        & Block-level ambient field vector.                                                                                      & $\mathbb{R}^q$              \\ \hline
		$\underline{\tilde{h}}$                            & Block-level control field vector.                                                                                      & $\mathbb{R}^q$              \\ \hline
		$G$                                                & Block membership matrix. $G_{ij}=1$ if node $i$ is in block $j$ and $G_{ij}=0$ otherwise.                              & $\mathbb{R}^{N \times q}$   \\ \hline
		$\underline{h}_{\mathrm{full}}$                    & Node-specific (full graph) Ising Influence strategy computed using Algorithm \ref{mf_iim_alg}.                         & $\mathbb{R}^N$              \\ \hline
		$\underline{h}_{\mathrm{block}}$                   & Block specific Ising influence strategy computed using the coarsening defined in Section \ref{block_level_approx_alg}. & $\mathbb{R}^N$              \\ \hline
		$M_{MC}(\underline{z})$                            & Average magnetisation estimated using Monte Carlo simulations for control field $\underline{z}$.                       & $[-1,1]$                    \\ \hline
		$\underline{h}_{\mathrm{unif}}$                    & Uniform influence strategy for which $h_{\mathrm{unif},i}=\frac{H}{N}$                                                 & $\mathbb{R}^N$              \\ \hline
		$\lambda_1$                                        & Largest eigenvalue of the adjacency matrix $A$.                                                                        & $\mathbb{R}$                \\ \hline
		$T_{burn}$                                         & Burn-in time for Monte Carlo simulations (see Section \ref{Ising_monte_carlo_sect}).                                   & $\mathbb{Z}+$               \\ \hline
		$T$                                                & Post burn-in run time for Monte Carlo simulations.                                                                     & $\mathbb{Z}+$               \\ \hline
		$\delta M^{Block}_F$                               & Fractional increase in mangetisation relative to uniform baseline (computed using Monte Carlo simulations).            & $\mathbb{R}^+$              \\ \hline
		$\underline{\hat{M}}(t)$                           & Estimate of the block level average magnetisation vector (elements $\hat{M}_{B_i}(t)$).                                & $[-1,1]$                    \\ \hline
		$\hat{\chi}_{snap}(t)$                             & Approximation of the susceptibility vector used in the snapshot influence strategy.                                    & $\mathbb{R}^q$              \\ \hline
		$\underline{h}_{snap}(t)$                          & Snapshot influence strategy.                                                                                           & $\mathbb{R}^N$              \\ \hline
		$H_{snap}$                                         & Sum of the elements of $ G \hat{\chi}_{snap}(t)$.                                                                      & $\mathbb{R}$                \\ \hline
		$\underline{m}^0$                                  & Initial condition used in fixed-point iteration of the full graph (Algorithm \ref{mf_mag_alg}).                        & $[-1,1]^N$                  \\ \hline
		$\underline{m}_B^0$                                & Initial condition used in fixed-point iteration applied to the coarsened graph (Algorithm \ref{mf_mag_alg}).           & $[-1,1]^q$                  \\ \hline
		$\epsilon$                                         & Step size in projected gradient ascent for Ising influence optimisation (Algorithm \ref{mf_iim_alg})                   & $\mathbb{R}+$               \\ \hline
		$a$                                                & Tolerance in Algorithm \ref{mf_iim_alg}.                                                                               & $\mathbb{R}+$               \\ \hline
		$\gamma$                                           & Damping term in the fixed-point iteration (Algorithm \ref{mf_mag_alg}).                                                & $\mathbb{R}+$               \\ \hline
		$\tau$                                             & Tolerance in the fixed-point iteration (Algorithm \ref{mf_mag_alg}).                                                   & $\mathbb{R}+$               \\ \hline
		$\psi$                                             & Threshold on the background field for hesitant targeting influence strategy.                                           & $\mathbb{R}+$               \\ \hline
		$\underline{h}_{HT}$                               & Hesitant targetting influence strategy.                                                                                & $\mathbb{R}^N$              \\ \hline
		$\underline{h}_{NC}$                               & Negative cancelling influence strategy.                                                                                & $\mathbb{R}^N$              \\ \hline
		$g$                                                & Linear field gradient acting on Pokec average age blocks (see Section \ref{Pokec_external_sim})                        & $\mathbb{R}$                \\ \hline
	\end{tabular}}
	\caption{Table of symbols used in the main text.}
	\label{symbols_table}
\end{table}

\section{Algorithms for mean-field Ising systems}

\subsection{Iterative schemes for solving the mean-field equations} \label{Mean_Field_Ising}

Equations \ref{ising_mf_equations} cannot be evaluated analytically, however, they can be solved using an iterative method. We will consider a fixed point iteration in which we update the magnetisation of each node according to \cite{aspelmeier2006free-energy,lynn2018maximizing}:
\begin{equation} \label{mf_iterative_dynamics}
m_i^{t+1} = (1-\gamma) m_i^t + \gamma \tanh\bigg( \beta \big( \sum_{j=1}^N A_{ij} m_j^t + g_i \big) \bigg)
\end{equation}
where we update the values in the order of their index (e.g. starting at $i=0$ finishing at $i=N$) (According to \cite{aspelmeier2006free-energy} the order in which we update for the different spins does not matter). This iterative method is sometimes referred to as \emph{damped fixed-point iteration} \cite{dahmen2016correlated,zou2011improved}. The damped fixed-point iteration scheme is summarised in Algorithm \ref{mf_mag_alg}.

\begin{center}
	\begin{minipage}{.8\linewidth}
		\begin{algorithm}[H]
			\SetAlgoLined
			\KwIn{Adjacency matrix $A$, inverse temperature $\beta$, external field vector $\underline{g}$, tolerance $\tau$, damping term $\gamma$, initial magnetisation $\underline{m}^0$.}
			\KwResult{Approximate solution to the mean-field self-consistency Equations \ref{ising_mf_equations} $\underline{m}^*$.}
			\textbf{Initialisation:} $t \leftarrow 0$, $i \leftarrow 0$ \;
			\While{$|\underline{m}^{t+1}-\underline{m}^t| > \tau $}{
				
				\While{$i<N$}{
					$m_i^{t+1} \leftarrow (1-\gamma) m_i^t + \gamma \tanh\bigg( \beta \big( \sum_{j=1}^N A_{ij} m_j^t + g_i \big) \bigg)$ ; \\
					$i++$ ; 
				}
				$t++$ ;
				$i \leftarrow 0$
			}
			\textbf{Return} $\underline{m}^{t+1}$
			\caption{Damped fixed point iteration}
			\label{mf_mag_alg}
		\end{algorithm}
	\end{minipage}
\end{center}

When solving the standard mean-field equations, standard fixed point (Picard) iteration \cite{berinde2007iterative} with $\gamma=1.0$ is sufficient in order secure convergence of the scheme \cite{aspelmeier2006free-energy}. However, when solving the self-consistency equations for higher order variational approximations, such as the TAP approximation, it becomes necessary to use the damped fixed point (Krasnoselskij) iteration in order to secure convergence \cite{aspelmeier2006free-energy,bolthausen2014iterative}.  Damped fixed point iteration allows us to obtain the fixed points of maps which satisfy less assumptions concerning their contractive properties (see theorem 3.2 in \cite{berinde2007iterative}).

For general Ising systems there will not be a unique solution to Equations \ref{ising_mf_equations}. The solution that we obtain using Algorithm \ref{mf_mag_alg} will depend on the state that we initialise the system in. Apart from when otherwise specified, we will focus on identifying the solution with the largest positive value of $M^{MF}$. For general $A$ and $\underline{g}$ the basins of convergence for the dynamical system defined in Equation \ref{mf_iterative_dynamics} will be difficult to identify. However, in practice we find that initialising $\underline{m}^0 = (1,1,...,1)$ results in convergence to the solution with largest positive $M^{MF}$.

\subsection{Control of Ising systems}

\subsubsection{Projected gradient ascent algorithm for Ising influence maximisation} \label{PGA_section}

We will estimate $\underline{h}^*$ by using a gradient ascent based approach. This standard approach for solving optimisation problems involves initialising the system at some initial state $\underline{h}^0$, and then making a series of steps in the direction of positive gradient of the objective function. In this case our aim is to maximise the magnetisation $M$. An accurate estimate of the gradient in the magnetisation could be obtained using Monte Carlo simulations, however, this would be costly as a new estimate of the gradient has to be made at each step. Consequently, we use the mean-field approximation to approximate the gradient. 

\textbf{The magnetisation gradient and susceptibility.} The gradient in the magnetisation with respect to the control field $\underline{h}$ is defined by:
\begin{equation} \label{Gradient_mag_def}
\nabla_{\underline{h}} M = \bigg(\frac{\partial M}{\partial h_1},\frac{\partial M}{\partial h_2},...,\frac{\partial M}{\partial h_N}  \bigg).
\end{equation}
The elements of this vector can be written in terms of the \emph{susceptibility matrix} $\chi$, which has elements: 
\begin{equation}
\chi_{ij} = \frac{\partial m_i}{\partial h_j} .
\end{equation}
Element $\chi_{ij}$ quantifies how changing the field experienced by node $j$ affects the magnetisation of node $i$. We will have:
\begin{equation}
\frac{\partial M}{\partial h_j} = \frac{\partial }{\partial h_j} \sum_{i=1}^N m_i = \sum_{i=1}^N \chi_{ij} . 
\end{equation}
Under the mean-field approximation the susceptibility can be obtained by differentiating Equation \ref{ising_mf_equations} in order to obtain:
\begin{equation} \label{mf_sus}
\chi^{MF} = \beta (I-\beta D A)^{-1} D ,
\end{equation}
where $D$ is an $N \times N$ matrix with elements $D_{ij}=(1-m_i^2) \delta_{ij}$. In Ref. \cite{lynn2016maximizing} $\chi^{MF}$ is obtained from Equation \ref{mf_sus}. This involves the inversion of an $N \times N$ matrix which becomes costly as the size of the graph increases. 

In Ref. \cite{tanaka1998mean-field} they show how, under the mean-field approximation the inverse of the susceptibility matrix is given by the matrix with elements:
\begin{equation} \label{stability_matrix}
S_{ij} = (\chi^{-1})_{ij} = \frac{1}{\beta} \frac{1}{1-m_i^2}\delta_{ij} - A_{ij}.
\end{equation}
The inverse of this matrix provides an estimate of the correlations under linear response theory. The first term diverges as $|m_i| \rightarrow 1$. This does not occur for the majority of the parameter regimes considered in the main text. However, we limit $\frac{1}{1-m_i^2}$ at $10^{20}$ in order to prevent the solver used in the step below from failing. In order to compute the gradient of the mean-field magnetisation we only need the elements of $\nabla_{\underline{h}} M^{MF}$ rather than all of the elements of the susceptibility matrix. The gradient can be written as:
\begin{equation} 
\nabla_{\underline{h}} M^{MF} = \chi^T \mathbf{1}, 
\end{equation}
where $\mathbf{1}=(1,1,...,1)$ is a vector of ones. Inverting and noting to swap the transpose and inverse we obtain:
\begin{equation} \label{linear_eq_for_sus}
S^{T} \nabla_{\underline{h}} M^{MF} =  \mathbf{1}.
\end{equation}
This linear equation can be solved using standard linear algebra routines which are more efficient than matrix inversion.

Given the gradient of the magnetisation at timestep $t$, we then make a step of magnitude: $ \epsilon \nabla_{\underline{h}} M^{MF} |_{\underline{h}^t}$ where $\epsilon$ is the step size parameter. Note that we normalise the gradient by its magnitude. The size of the step which we take could vary with iteration. Various algorithms including `line search' based approaches exist for estimating the optimum step size to take at each iteration during optimisation procedures \cite{nocedal2006numerical}. For simplicity we will consider the case of fixed step size. We find that we can obtain convergence in a feasible number of iterations with fixed step size in practice for the parameters considered in this paper.

\textbf{Satisfying the budget constraint. } If the proposed $\underline{h}^{t+1}$ does not satisfy the imposed budget constraint (Equation \ref{budget_constraint}) we identify a projection of this vector which does. A naive implementation of this step can be very slow. Consequently, it is of interest to find more efficient algorithms for carrying out this projection. In order to improve the performance we use an implementation of the methods discussed in Ref. \cite{blondel2014large-scale}. The code used for this step is available at: \href{https://gist.github.com/mblondel/6f3b7aaad90606b98f71}{https://gist.github.com/mblondel/6f3b7aaad90606b98f71}. They discuss how projections onto the $l_1$-ball can be treated using methods for performing projections onto the simplex. These methods were found to be relatively efficient compared to a naive implementation.

\textbf{Description of the algorithm. } Combining the steps described above we obtain an update rule for the control field:
\begin{equation}
\underline{h}^{t+1} = \mathcal{P}_{|\underline{h}| \leq H} \bigg[ \underline{h}^t + \epsilon \nabla_{\underline{h}} M^{MF}  \bigg], 
\end{equation}
where $\mathcal{P}_{|\underline{h}| \leq H}\big[ \cdot \big]$ represents the projection into the set of fields which satisfy the budget constraint. We can repeatedly update according to this rule until the change in the magnetisation is less than some threshold parameter $a$. The resulting procedure for determining the control field is shown in Algorithm \ref{mf_iim_alg}. At each step of Algorithm \ref{mf_iim_alg} we use Algorithm \ref{mf_mag_alg} to compute $M^{MF}(\underline{b}+\underline{h}^{t})$.

\begin{center}
	\begin{minipage}{.8\linewidth}
		\begin{algorithm}[H]
			\SetAlgoLined
			\KwIn{Adjacency matrix $A$, ambient field $\underline{b}$, step size $\epsilon$, tolerance $a$, inverse temperature $\beta$, initial control field $\underline{h}^0$.}
			\KwResult{Estimate of the optimal control field $\underline{h}^*$.}
			
			Initialisation $t \leftarrow 0$ ;
			
			\While{$|M^{MF}(\underline{b}+\underline{h}^{t})-M^{MF}(\underline{b}+\underline{h}^{t-1})| > a$}{
				
				$\underline{h}^{t+1} = \mathcal{P}_{|\underline{h}| \leq H} \bigg[ \underline{h}^t + \epsilon \nabla_{\underline{h}} M^{MF}  \bigg]$;
				
				$t++$;
			}
			\textbf{Return} $\underline{h}^{t+1}$
			\caption{Projected gradient ascent}
			\label{mf_iim_alg}
		\end{algorithm}
	\end{minipage}
\end{center}

This algorithm is the same as that in \cite{lynn2016maximizing}. However, we found that computing the susceptibility from Equation \ref{linear_eq_for_sus}, rather than Equation \ref{mf_sus} resulted in a significant speed-up. This, combined with the efficient method used for computing the projection, was found to result in 
a significant overall speed-up. Numerical simulations suggest that the improvement can be several orders of magnitude. This speed-up meant that we can relatively quickly obtain optimal fields for graphs containing a few thousand nodes. For much larger graphs this algorithm will still scale poorly. As a result, we now move on to describe methods which make use of the modular structure which might be inherent in the system.

\subsubsection{Block level approximation for the susceptibility} \label{block_level_approx_alg}

We now demonstrate how, under assumptions of block structure, we can reduce the system of $N$ coupled equations for the mean-field magnetisation down to $q$ equations at the level of blocks. After averaging over the disorder in the block structure, the treatment here follows a similar line of that of the multi-species Curie-Weiss model (e.g. see Refs. \cite{fedele2017inverse,opoku2019parameter}). Recall that the mean-field self-consistency equations are given by:
\begin{equation}
m_i = \tanh\bigg( \beta \big( \sum_{j=1}^N A_{ij} m_j + g_i \big)  \bigg),
\end{equation}
for $i=1,...,N$. We will assume that the adjacency matrix is drawn from a network with an SBM structure as in Section \ref{SBM_section} with $q$ blocks. We define the average magnetisation of block $k$ to be:
\begin{equation}
m_{B_k} = \frac{1}{N_k} \sum_{k \in \mathcal{C}_k} m_k,
\end{equation}
where, $\mathcal{C}_k$ is the set of nodes in block $k$.

We now make the assumption that we can average over the disorder in the network structure by replacing the adjacency matrix with the expected adjacency matrix (see Ref. \cite{lowe2019multi-group}). In this case, the mean-field self-consistency equations become:
\begin{equation} \label{mf_with_average_A}
m_i = \tanh\bigg( \beta \big( \sum_{j=1}^N \mathbb{E}(A)_{ij} m_j + g_i \big)  \bigg).
\end{equation}
In this system of equations the couplings experienced by nodes in the same block will be identical. We further assume that all nodes in the same block $k$ will experience the same total external field $\tilde{g}_{B_k}$. The total field applied to the system can be characterised by the vector of length $q$: $\tilde{\underline{g}} = (\tilde{g}_{B_1},...,\tilde{g}_{B_q})$. This vector can be projected onto the full system of size $N$ by multiplying by the block membership matrix to obtain:
\begin{equation}
\underline{g}_B = G \tilde{\underline{g}} , 
\end{equation}
where $G$ is the block membership matrix for which $G_{ij}=1$ if node $i$ is in block $j$ and $G_{ij}=0$ otherwise. The block-level total external field can be decomposed into the block-level ambient field $\tilde{\underline{b}}$ and the block-level control field $\tilde{\underline{h}}$ so that $\tilde{\underline{g}}=\tilde{\underline{b}}+\tilde{\underline{h}}$. When performing the optimisation at the level of blocks the budget constraint can be written as:
\begin{equation}
\sum_{k=1}^q N_k \tilde{h}_{B_k} = H .
\end{equation}

Under these assumptions the magnetisations of all nodes in block $k$ will be equal to $m_{B_k}$. For a node $i$ in block $k$ the first term in the brackets in Equation \ref{mf_with_average_A} can be written as:
\begin{equation}
\sum_{j = 1}^N \mathbb{E}(A)_{ij} m_j = (N_k-1) \Omega_{kk} m_{B_k} + \sum_{j \ne k} N_j \Omega_{kj} m_{B_j} \\
\approx \sum_{j=1}^q N_j \Omega_{kj} m_{B_j}
\end{equation}

Using the definition of the coupling matrix (Equation \ref{sbm_coupling_matrix}) we can write the original set of $N$ coupled equations for the magnetisations of the nodes as a set of $q$ equations describing the average magnetisations at the level of blocks:
\begin{equation} \label{mf_equations_block_level_app}
m_{B_x} = \tanh\bigg( \beta \big(\sum_{y=1}^q \mathcal{K}_{xy} m_{B_y} +  \tilde{g}_{B_x} \big) \bigg),
\end{equation}
for $x=1,...,q$. The \emph{total magnetisation}, which we wish to maximise, can be written as:
\begin{equation}
M^{MF} = \sum_{x=1}^q N_x m_{B_x} . 
\end{equation}
The gradient in the magnetisation with respect to the block-level control will have elements:
\begin{equation}
\big( \nabla_{\tilde{\underline{h}}_B} M^{MF} \big)_y = \frac{\partial}{\partial \tilde{h}_{B_y}} M^{MF} = \sum_{w=1}^{q} N_w \tilde{\chi}_{wy}^{MF}  , 
\end{equation}
where $\tilde{\chi}^{MF}$ is the $q \times q$ block-level mean-field susceptibility matrix. This matrix will have elements:
\begin{equation}
\tilde{\chi}_{xy}^{MF} = \frac{\partial m_{B_x}}{\partial \tilde{h}_{B_y}} = \beta (1-m_{B_x}^2) \bigg( \sum_{p=1}^q N_p \Omega_{xp} \frac{\partial m_{B_p}}{\partial \tilde{h}_{B_y}} + \delta_{xy} \bigg)
\end{equation}
Defining $D_M$ such that $D_{M_{xy}} = (1-m_{B_x}^2) \delta_{xy}$ allows us to write an expression for the block-level susceptibility matrix:
\begin{equation}
\tilde{\chi}^{MF} = \beta D_M (I + \mathcal{N} \Omega \tilde{\chi}^{MF} )
\end{equation}
Consequently, the analogue of the stability matrix (Equation \ref{stability_matrix}) for the block-level system $\tilde{S}$ will therefore have the elements:
\begin{equation}
\tilde{S}_{xy} = \big( \tilde{\chi}^{MF^{-1}} \big)_{xy} = \frac{1}{\beta} \frac{1}{1-m_{B_x}^2} - N_x \Omega_{xy} . 
\end{equation}
We can identify the magnetisation gradient at the level of blocks by solving:
\begin{equation} \label{block_mf_sus_equation}
\tilde{S} \nabla_{\tilde{\underline{h}}_B} M^{MF} =  \mathbf{1}.
\end{equation}

\subsection{Monte Carlo simulations} \label{Ising_monte_carlo_sect}

The standard method to obtain samples from intractable probability distributions, such as $\mathbb{P}(\underline{s})$ is via Monte Carlo simulations \cite{newman1999monte}. The \emph{Markov chain Monte Carlo} method involves drawing a series of samples $\underline{s}_1,\underline{s}_2,\underline{s}_3,...$ from $\mathbb{P}(\underline{s})$ by constructing a Markov chain which has $\mathbb{P}(\underline{s})$ as a stationary distribution.

Given some current state of the system $\underline{s}$ there are many ways to choose a next state $\underline{s}'$. A simple choice is \emph{single spin flip dynamics} for which we select a spin at random to flip at each timestep. We use the \emph{Methopolis Hastings} acceptance probability, for which the probability of transitioning from $\underline{s}$ to $\underline{s}'$ at a given time step in the dynamics is given by:
\begin{equation} \label{acceptance_prob}
A(\underline{s} \rightarrow \underline{s}') = \begin{cases}
e^{-\beta \Delta E} \quad \mathrm{if} \quad \Delta E > 0 \\
1 \quad \mathrm{otherwise}
\end{cases},
\end{equation}
where the change in energy is given by:
\begin{equation} \label{MC_energy_change}
\Delta E = 2 s_i \bigg( \sum_{j \in \mathcal{N}(i)} s_j + g_i \bigg),
\end{equation}
where $\mathcal{N}(i)$ represents the set of neighbours of node $i$. This is a generalisation of Equation (3.10) from Ref. \cite{newman1999monte} which also includes the external field $g_i$. This is the most efficient choice of the acceptance probability \cite{newman1999monte}. The intuition behind this choice is that we accept any spin flip which lowers the energy of the system. For flips which increase the energy of the system we accept the flip with a probability which decays exponentially with the increase in energy.

\subsection{Setting the temperature scale}  \label{crit_temp_discuss}

In the main text we consider influencing Ising systems both above and below the phase transition. For the purposes of this paper it is not necessary to estimate the critical temperature precisely, however, it is relevant to obtain a guideline value using the mean-field approximation. We estimate this baseline for the case of zero ambient field.

For the purpose of this analysis we treat the mean-field equations as a discrete time dynamical system. In the presence of no external fields ($\underline{g}=0$) the mean-field equations take the form:
\begin{equation} \label{mf_w_no_field}
m_i^{t+1} = \tanh\bigg( \beta \big( \sum_{j=1}^N A_{ij} m_j^t  \big)  \bigg) = \mathcal{F}_i(\underline{m}^t),
\end{equation}
where $\mathcal{F}_i(\cdot)$ represents the $ith$ component of the map. This system always has a fixed point at $\underline{m}=0$.

Following Ref. \cite{lynn2016maximizing} we identify the critical temperature by identifying the conditions under which the fixed point at $\underline{m}=0$ is the only fixed point. The contraction mapping theorem states that a fixed point is unique if a map is a contraction \cite{brooks2009contraction}. Consequently, if we can identify conditions under which Equation \ref{mf_w_no_field} is not a contraction then the fixed point is necessarily not unique.

We can analyse the stability of a fixed point by considering the Jacobian matrix $J$, which has elements $J_{ik} = \frac{\partial \mathcal{F}_i}{\partial m_k}$. For Equation \ref{mf_w_no_field} the elements of the Jacobian matrix take the form:
\begin{equation}
J_{ik} = \beta \bigg( 1 - \tanh^2\bigg( \beta \big( \sum_{j=1}^N A_{ij} m_j  \big)  \bigg) \bigg) A_{ik} = \beta (1-m_i^2) A_{ik} .
\end{equation}
The eigenvalues of $J$ evaluated at $\underline{m}=0$ take the form $\beta \lambda_i$, where $\lambda_1 \leq \lambda_2 \leq ... \leq \lambda_N$ are the eigenvalues of the $A$. If any of the eigenvalues of $J$ are greater than one then the fixed point will be unstable along the direction associated with the corresponding eigenvector. For the fixed point at $\underline{m}=0$ this will be the case if $\beta \lambda_N > 1$. Increasing $\beta$ sufficiently to meet this condition will result in a bifurcation. During this bifurcation the fixed point at $\underline{m}=0$ becomes unstable and we observe the formation of two new fixed points. If the largest eigenvalue is degenerate then it is also possible for more than two fixed points to form during the bifurcation.

We therefore define the \emph{critical temperature} of the system by:
\begin{equation} \label{spectral_equation}
\beta_c = \frac{1}{\lambda_N}.
\end{equation}
This formula applies when $\underline{g}=0$, however, we find in practice that $\beta_c$ gives a good guideline value for the interface between the high and low temperature regimes when ambient fields are present.

\subsection{Parameters used in Figure 1} \label{Parameters_two_block}

In order to generate the results for Figure \ref{two_block_markups} we implement projected gradient ascent (Algorithm \ref{mf_iim_alg}) at the level of the full graph and at the level of blocks. We use step sizes of $\epsilon=50$ and $1$ and tolerances of $a=10^{-6}$ and $a=10^{-8}$ for the full and block-level IIM respectively. In each instance of the above algorithms we solve for the mean-field magnetisation using Algorithm \ref{mf_mag_alg} with $\gamma= 1$ and tolerance of $ \tau = 10^{-5}$.

\section{Impact of degree heterogeneity on coarse-grained influence} \label{degree_hetero_impact}

In section \ref{ising_scalable} in the main text we consider coarse-grained influence on a two-block SBM with heterogeneous blocks. In this section we consider the impact of the average block degree on our ability to influence the system. We consider a system with a connectivity matrix of the form:
\begin{equation}  \label{two_block_coupling_matrix}
\mathcal{K} = \begin{pmatrix}
\mathcal{K}_{11} & 2.5 \\
2.5 & 5.5
\end{pmatrix},
\end{equation}
where tuning $\mathcal{K}_{11}$ allows us to tune the relative average degree of the two blocks.

\begin{figure*}
	\centering
	\subfloat[]{\label{kap_imp_bet05}\includegraphics[width=.32\textwidth]{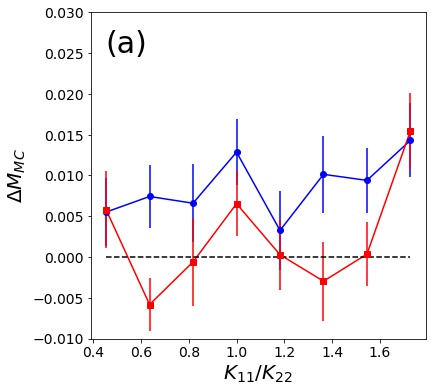}}
	\hspace*{0.6em}
	\subfloat[]{\label{kap_imp_bet12}\includegraphics[width=.32\textwidth]{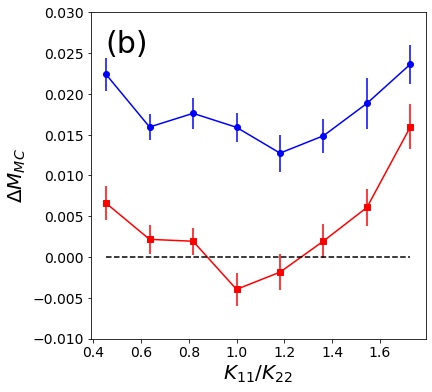}}
	\hspace*{0.6em}
	\subfloat[]{\label{kap_imp_bet15}\includegraphics[width=.32\textwidth]{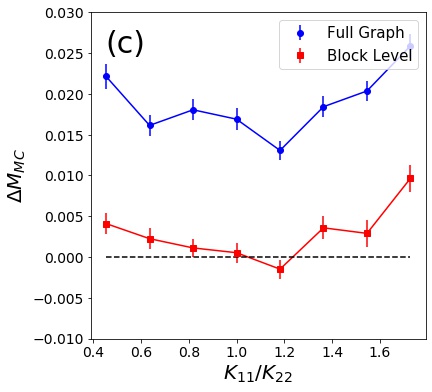}}
	\vspace*{-1.0em}
	\caption{\textbf{Coarse-grained influence strategies perform best when average block degrees are heterogeneous.} Plot showing how $\Delta M(\protect\underline{h}_{\mathrm{IIM}})$ (blue circles) and $ \Delta M(\protect\underline{h}_{\mathrm{block}})$ (red squares) behave as a function of $K_{11}/K_{22}$ for $\beta = 0.5 \beta_c$ (a), $1.2 \beta_c$ (b) and $ 1.5 \beta_c$ (c). Shown for the case of $H=2000$ with the other parameters the same as those in main text Figure \ref{two_block_markups}. In each case we show the average of 15 Monte Carlo simulations with the error bars representing the standard error on the mean. Parameters used in the optimisation and Monte Carlo simulations are the same as those in Figure \ref{two_block_markups}. In both cases the magnetisation markup is largest when the average degrees of the blocks vary. However, the coarse-grained approximation can only achieve a significant advantage when the average degrees are relatively heterogeneous.}
	\label{kappa_impact}
\end{figure*}

We compute the magnetisation markup for both the full-graph mean-field and block-level influence strategies for $H=2000$ for $\beta = 0.5 \beta_c $, $1.2 \beta_c$ and $ 1.5 \beta_c$ respectively. This field budget was chosen as it represents the one where we saw a relatively large markup in Figure \ref{two_block_markups}. The impact of varying the average degree at different temperatures is illustrated in Figure \ref{kappa_impact}. We confirm that, as we might expect, the coarse-grained strategy in this system is most effective when there is sufficient variability in the average degrees of the coarse-grained blocks.

\section{Initial conditions for computing phase diagrams} \label{three_block_phase_diag}

Ising systems consisting of weakly connected groups of spins can possess metastable states where the groups are polarised in a certain direction. For the three-block system considered in Section \ref{three_block_sbm_section} we can imagine $2^3 = 8$ different metastable states occurring. The addition of ambient external fields will break the symmetry of the system. This will result in states in which the spins are aligned with the ambient fields occurring with a higher probability. In order to determine the full phase diagram of the system, we must take these metastable states into account. We will now describe how we achieve this for the mean-field approximation and the Monte Carlo simulations.

\textbf{Initialisation of block-level fixed point iteration. } For the block-level approximation on the three-block system described above $\underline{m} \in [1,1]^3$. Initialising Algorithm \ref{mf_mag_alg} at different initial states when computing the mean-field magnetisation will result in convergence to different solutions. As noted previously in Section \ref{Mean_Field_Ising}, it is not easy to determine the basins of attraction for the convergence of the Algorithm \ref{mf_mag_alg}. Consequently, we identify different solutions by sampling \emph{uniform} random initial conditions within the domain. In order for this approach to work, we have assumed that the basins of attraction for the different metastable states are sufficiently large that we can reach them after a moderate number of random samples. We use 100 random initial conditions for each value of $\beta$ when making Figure \ref{three_block_phase}.

\textbf{Initialisation of full graph fixed point iteration. } When solving the mean-field self-consistency equations for the full graph we find that using uniformly random initial conditions in $[-1,1]^N$ identifies relatively few metastable states. This may be because certain metastable states, such as those close to being fully aligned, have basins of attraction in the far corners of the high-dimensional hypercube $[-1,1]^N$. Consequently, we instead randomly sample initial states of the form:
\begin{equation} \label{mf_three_block_initial}
\underline{m}_0' = (a_1 \mathbf{1},a_2 \mathbf{1}, a_3 \mathbf{1}),
\end{equation}
where $a_i \sim U[-1,1]$. A particular random draw of $a_i$ will set the magnetisation of all nodes in block $i$ to be $a_i$. This sampling strategy assumes that all the nodes in a given block have approximately the same average magnetisation which is a reasonable assumption for weakly coupled modular graphs. We used 16 random initial conditions for each value of $\beta$ when making Figure \ref{three_block_phase}.

\textbf{Initialisation of Monte Carlo simulations. } Choosing to initialise Monte Carlo simulations randomly (i.e. with +1/-1) will lead to an initial state with $M \approx 0$. Starting from such an initial condition, it is more likely for the system to come to a temporary equilibrium in a more disordered metastable state. This makes it difficult to identify the more aligned metastable states (for values of $\beta$ where they exist). Consequently, rather than waiting a long time for the system to transition between different metastable states we instead use an analogue of Equation \ref{mf_three_block_initial}. We will consider an initial condition of the form:
\begin{equation}
\underline{m}_{\mathrm{MC}} = (\underline{\eta}(a_1),\underline{\eta}(a_2),\underline{\eta}(a_3)),
\end{equation} 
where $\underline{\eta}(a_i)$ is a vector of length $N_i$ for which $\eta(a_i)_i = +1$ with probability $a_i$ and  $\eta(a_i)_i = -1$ with probability $1-a_i$.

In our simulations we combine samples from multiple Monte Carlo chains after burn-in. We then bin the sampled magnetisation values into equal sized bins in $[-1,1]$. This allows us to build up a picture of the distribution of magnetisations for each value of $\beta$.

The solution to the full graph mean-field approximation shown in Figure \ref{three_block_phase} lacks symmetry at the first bifurcation point. We have not fully investigated the cause of this. We expect that the particular SBM draw has an imbalance in the number of connections between the edge blocks and the centre block. This will mean that one of the two metastable states will become stable marginally earlier than the other. We expect that it would also be possible to identify this using Monte Carlo simulations, however, it is difficult to see this effect due to noise given the system size considered. We expect such asymmetries in the solution to the full graph mean-field approximation might become diminished for larger system sizes.

\subsection{Impact of ensemble variability on phase diagram} \label{ens_var_phase_diag}

\begin{figure*}
	\centering
	\subfloat[]{\includegraphics[width=.6\textwidth]{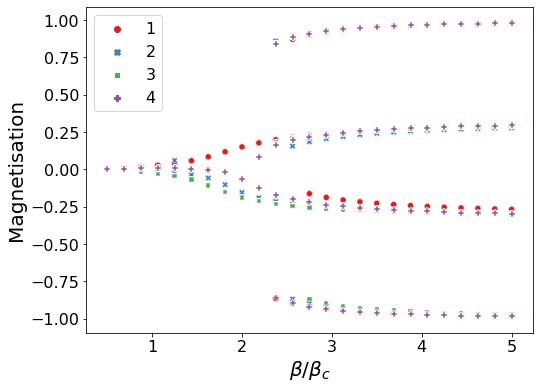}}
	\vspace*{-2.0em}
	\caption{\textbf{Impact of ensemble variability on the magnetisation for a three block SBM}. Plot showing $M^{MF}$ as a function of $\beta$ for different SBMs drawn from the ensemble described in Section \ref{three_block_sbm_section}. For each value of $\beta$ for each graph we run Algorithm \ref{mf_mag_alg} 50 times with different random initial conditions selected according to the scheme described in Supplementary Section \ref{three_block_phase_diag} above. }
	\label{three_block_ensemble}
\end{figure*}

In Figure \ref{three_block_phase} in the main text we computed the magnetisation for a single realisation of the SBM drawn from the ensemble described in Section \ref{three_block_sbm_section}. The phase diagram showed a lack of symmetry about the bifurcation point. Figure \ref{three_block_ensemble} shows the phase diagram for four different SBMs drawn from the ensemble. We see that the different realisations are asymmetric, however, their average is symmetric. We would intuitively expect that as $N \rightarrow \infty$ that the ensemble average value of $M^{MF}$ for different realisations of the full graph will converge to the symmetric phase diagram for the block-level approximation shown in Figure \ref{three_block_phase}. However, this convergence might be based on certain constraints on the system such as sufficient homophily and density (see Theorem 3.1 in Ref. \cite{lowe2019multi-group}).

\section{Data processing and simulation methodology associated with the Pokec social network}

\subsection{Pokec Dataset} \label{pokec_data_description}

In order to demonstrate the applicability of the algorithm we will consider the IIM problem on a real world social network. We use a dataset obtained from the online social network Pokec \cite{takac2012data}. This dataset was access from: \href{https://snap.stanford.edu/data/soc-Pokec.html}{https://snap.stanford.edu/data/soc-Pokec.html}. The Pokec network contains $N=1,632,803$ attributed nodes. The network data is directed but we will treat the edges as undirected.

The 58 different attributes consist of a range of numerical and text fields, a significant fraction of which are incomplete. We will focus on age and location as these attributes are relatively well populated compared to the other attributes. Furthermore, we also expect to see homophily with respect to these two attributes meaning that we can classify nodes into blocks without having to rely on knowledge of the graph structure in order to determine the communities.

\textbf{Obtaining spatial coordinates. } The locations of users are not stored as geographic coordinates. Instead, we have the names of the region or towns that the users are based at. These are in Slovak and stored in the column named `region'. In order to convert the region names to geographic coordinates we use one of the geocoders from the geopy package (see: \href{https://pypi.org/project/geopy/}{https://pypi.org/project/geopy/}). This allows us to extract the latitude longitude coordinates for all of the place names. A small percentage of the queries fail due to unrecognised place names. In addition, there also exist users with the term `zahranicie' which translates to abroad. We recover approximate latitude longitude coordinates associated with these users, however, we will not consider these profiles in the analysis.

After the above data processing we are left with 188 unique location names. After keeping valid latitude longitude coordinates and ages we are left with 64\% ($N=1042110$)  of the original nodes.


\subsection{Coarse graining using node demographics} \label{Pokec_block_extract}

We will consider a smaller subgraph of Pokec with supposed block structure. We consider a subset of the regions around Bratislava in the Pokec dataset as this domain contains several regions with a large population which are relatively well connected. Starting from the set of valid profiles described above we extract a large connected subgraph from the data as follows:
\begin{enumerate}
	\item{Select the set of profiles with region names containing the string `bratislava'.}
	\item{We then select the 3 regions with the largest population: 
		\begin{enumerate}
			\item{`bratislavsky kraj, bratislava - petrzalka'}
			\item{`bratislavsky kraj, bratislava - ruzinov'}
			\item{`bratislavsky kraj, bratislava - okolie'}
		\end{enumerate}
		Petrzalka and Ruzinov correspond to two densely populated regions on the right hand side and left hand side of the river, while `Okolie' refers to the surrounding districts.}
	\item{We take the subgraph of users within these regions and then extract the LCC. The motivation for focusing on the LCC is that disconnected nodes will be unaffected by the optimisation on the graph structure.}
	\item{The age distribution of Pokec users is skewed towards relatively young users. The 10th and 90th percentiles lie at ages 14 and 37 respectively. In order to obtain approximately equal sized age groups we will rely on percentiles rather than age categories of equal size. We consider age bins with boundaries at the 25th,50th and 75th percentiles. Using the age distribution of all users containing ages this corresponds the bins:
		\begin{enumerate}
			\item{1-17}
			\item{18-21}
			\item{22-28}
			\item{29-112}
	\end{enumerate}}
\end{enumerate}
Some of the larger ages in the dataset appear to be incorrectly reported. However, this only impacts a small number of the nodes so will not substantially skew the results. The resulting connected undirected subgraph from Pokec contains $N=29582$ users divided into 12 different blocks. The sizes of the different blocks are reported in table \ref{pokec_block_size_table}. Counting the edges between the blocks and inverting Equations \ref{edges_from_con_prob} and \ref{sbm_coupling_matrix} allows us to obtain the coupling matrix, $\mathcal{K}$, for the system. Figure \ref{pokec_interactions} shows a heatmap of the coupling strengths between the different blocks extracted. The diagonal element is largest indicating the presence of attribute homophily. We also observe larger off-diagonal elements which indicate the presence of homophily between age categories in different regions. We also visualise the coupling between the different groups in Figure \ref{social_connects}.

\subsection{Ambient field} \label{Pokec_external_sim}

\begin{figure*}
	\centering
	
	\subfloat[]{\label{social_connects}\includegraphics[width=.3\textwidth]{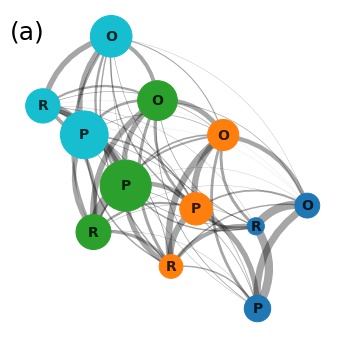}}
	\hspace*{8.0em}
	\subfloat[]{\label{linear_field}\includegraphics[width=.3\textwidth]{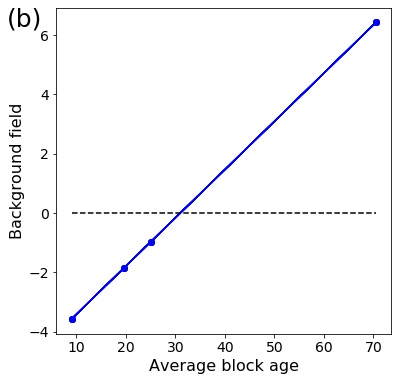}}
	\vspace*{-2.0em}
	\caption{\textbf{Homophily and ambient field set-up in the coarse-grained network.} (a) Network plot illustrating the connectivity between different groups in the Pokec social network (P = `Petrzalka', R=`Ruzinov' \& O =`Okolie'). Edge widths are proportional to the elements of the courpling matrix $\mathcal{K}$ (see Figure \ref{pokec_interactions}). Nodes are coloured according to the age category (blue: 1-17, orange: 18-21, green: 22-28 and turquoise: 29-112) and have sizes proportional to the size of each group (see Table \ref{pokec_block_size_table}). (b) Linear ambient field applied to the blocks for the case of $g=10$. The three blocks with lower ages have a small negative field, while the final has a strong positive ambient field.}
	\label{pokec_linear_field_and_connects}
\end{figure*}

Let $\mathcal{A}_i$ be the average age of the individuals in block $i$ (we use the average of the original age bins before diving into regions for simplicity). We first standardise the ages so that they have a mean of zero and lie in the range $[-1,1]$. Let: 
\begin{equation}
\hat{\mathcal{A}}_i = \frac{\mathcal{A}_i - \langle \mathcal{A}_i \rangle}{\max \mathcal{A}_i - \min \mathcal{A}_i}, 
\end{equation}
where $\langle \mathcal{A}_i \rangle$ is the average age of block $i$. We then an consider ambient field on block $i$ be given by:
\begin{equation} \label{pokec_age_fields}
\mathcal{B}_{\mathrm{age},i} = g \hat{\mathcal{A}}_i,
\end{equation}
where $g$ is a parameter controlling the field gradient. Ambient fields for the Pokec dataset for $g=10$ are shown in Figure \ref{linear_field}.


\begin{table}[]
	\resizebox{.3\textwidth}{!}{%
		\begin{tabular}{|
				>{\columncolor[HTML]{FFFFFF}}r |
				>{\columncolor[HTML]{FFFFFF}}r |}
			\hline
			\textbf{Block}          & \textbf{Size} \\ \hline
			Okolie ages 1-17      & 1234          \\ \hline
			Okolie ages 18-21     & 1939          \\ \hline
			Okolie ages 22-28     & 3154          \\ \hline
			Okolie ages 29-112    & 3458          \\ \hline
			Petrzalka ages 1-17   & 1398          \\ \hline
			Petrzalka ages 18-21  & 2106          \\ \hline
			Petrzalka ages 22-28  & 5234          \\ \hline
			Petrzalka ages 29-112 & 4550          \\ \hline
			Ruzinov ages 1-17     & 586           \\ \hline
			Ruzinov ages 18-21    & 1123          \\ \hline
			Ruzinov ages 22-28    & 2444          \\ \hline
			Ruzinov ages 29-112   & 2356          \\ \hline
		\end{tabular}%
	}
	\label{pokec_block_size_table}
	\caption{Number of nodes within each of the blocks defined based on individual's ages and regions.}
\end{table}

\subsection{Phase diagram} \label{Pokec_phase_diagram}

Before comparing influence strategies on the Pokec social network it is useful to explore the circumstances in which the coarse-grained approximation of the network allows us to accurately estimate the magnetisation of the system. We computed the phase diagram of the system for three different values of the field gradient $g$. This was performed using Monte Carlo simulations, the full mean-field approximation and the block-level mean field approximation. In the first two cases we initialise the system only at the fully negative (-1,-1,...,-1) and fully aligned (1,1,...,1) with the goal of identifying the most positive and most negative metastable states. In the case of the block-level approximation we sample different random initial conditions for each value of $\beta$ in a manner comparable to that described in section \ref{three_block_phase_diag}. 

\begin{figure*}
	\centering
	
	\subfloat[]{\label{pokec_pase_grad_1}\includegraphics[width=.32\textwidth]{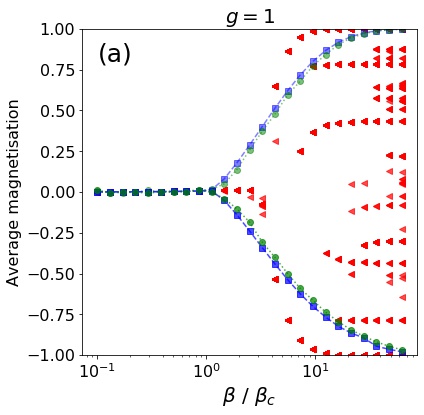}}
	\subfloat[]{\label{pokec_phase_grad_5}\includegraphics[width=.32\textwidth]{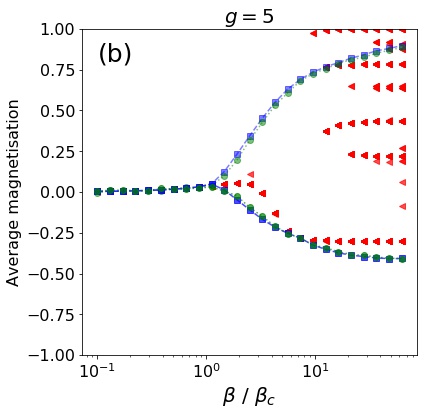}}
	\subfloat[]{\label{pokec_phase_grad_10}\includegraphics[width=.32\textwidth]{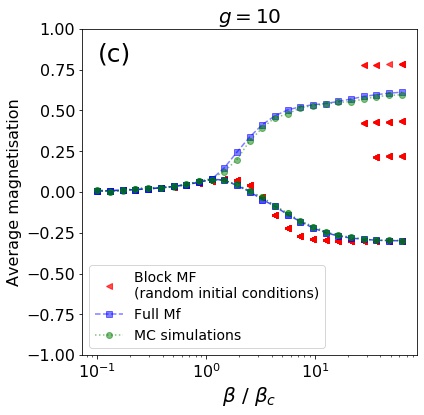}}
	\vspace*{-2.0em}
	\caption{\textbf{The magnetisation of the Pokec social network can be estimated using a coarse-grained approximation.} Phase diagram for the Pokec social network computed using Monte Carlo simulations (green circles), mean-field approximation on the full graph (blue squares) and the mean-field approximation on the coarse-grained system (red triangles). Shown for different values of the linear field gradient (a) $g=1$, (b) $g=5$ and (c) $g=10$. For the full graph mean-field approximation and Monte Carlo simulations we only consider the most positive and negative solution branches. Different metastable solutions were obtained for the block-level approximation by sampling 100 random initial conditions in the fixed point iteration (Algorithm \ref{mf_mag_alg}) for each value of $\beta$. Fixed point iteration was carried out with $\gamma=1$, $\tau=10^{-5}$ and a max number of iterations of $10^4$ for both the full graph and block-level approximation. For each value of $\beta$ we evaluate the average magnetisation using a single Monte Carlo chain with $T = 10^4$ and a burn-in time of $T_{\mathrm{burn}} = 3 \times 10^5$. We perform the simulations for 25 equally spaced (on a log base 10 scale) values of $\beta / \beta_c$ varying between $10^{-1}$ and $10^{1.8}$, where $\beta_c$ is obtained from the spectral radius of the adjacency matrix of the full graph.}
	\label{pokec_phase_diagram}
\end{figure*}

Figure \ref{pokec_phase_diagram} shows the phase diagram of the system for three different values of the field gradient $g$. The block-level system was found to have a numerous metastable states at larger values of $\beta$. The different metastable solutions correspond to solutions with different blocks being aligned and anti-aligned with each other. These solutions will not exactly match those in the full system as the partitions are relatively coarse and based on attributes rather than being inferred from any observed modular structure in the network. Nonetheless, the behaviour of the average magnetisation as a function of $\beta$ is broadly consistent with that obtained using the full graph mean-field approximation and Monte Carlo simulations. As the field gradient is increased we observe that some of the positive metastable solutions become suppressed and exist for a smaller range of $\beta$ values. This results from 3 out of 4 of the age groups having a negative ambient field (see Figure \ref{linear_field}).

\subsection{Impact of sample fraction on survey-snapshot control} \label{snapshot_convergence}

The survey-snapshot control introduced in Section \ref{three_block_structure} is derived from a snapshot of the Ising system's state at a particular point in time. In this section we explore the consequences of estimating $\hat{M}_{B_x}$ using a sample of $\lceil F_{\mathrm{samp}} N_{B_x}  \rceil$ nodes (sampled without replacement) from block $x$. The fraction of nodes sampled within each block will impact how well we can estimate the magnetisation of the system and therefore how well we can influence the system. This analysis assumes that a random sample is equivalent to a representative sample of the population, which is not necessarily the case in practice.

For a particular snapshot, we can quantify the quality of our estimate of the average magnetisation of a block $B_i$ for a particular value of $F_{\mathrm{samp}}$ (the fraction of nodes sampled) by computing the absolute deviation from the average magnetisation. We can quantify this deviation with the quantity:
\begin{equation}
\delta \hat{M}_{B_x} (F_{\mathrm{samp}}) = | \hat{M}_{B_x}(F_{\mathrm{samp}}) - \langle \hat{M}_{B_x}(1) \rangle |, 
\end{equation}
where $\langle \cdot \rangle$ indicates the average over multiple samples. Averaging this quantity over blocks allows us to obtain an overall picture of how accurately we can estimate the magnetisation for a particular fraction of nodes sampled. There are two sources of variance in $\hat{M}_{B_x}(F_{\mathrm{samp}})$: that due to the sub-sample of nodes and that due to the particular snapshot in time. Averaging $\delta \hat{M}_{B_x} (F_{\mathrm{samp}})$ over multiple draws allows us to assess the impact of the former.

\begin{figure*}
	\centering
	
	\subfloat[]{\label{snapshot_mags}\includegraphics[width=.4\textwidth]{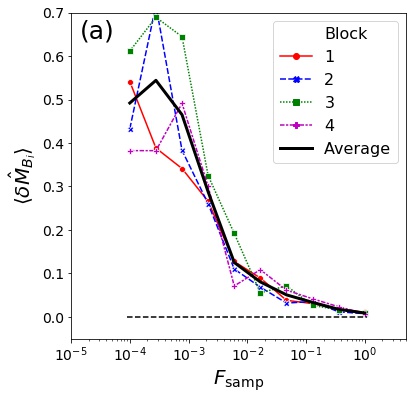}}
	\hspace*{4.0em}
	\subfloat[]{\label{snapshot_cons}\includegraphics[width=.4\textwidth]{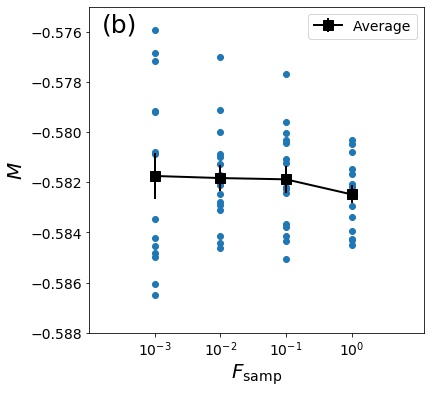}}
	\vspace*{-2.0em}
	\caption{\textbf{The average performance of the survey-snapshot influence strategy is unaffected by sample size, however, consistency increases with fraction of nodes sampled.} (a) Plot showing how $\langle \delta \hat{M}_{B_i} \rangle$ (averaged over 10 realisations of $\hat{M}_{B_i}(F_{\mathrm{samp}})$) varies as a function of $F_{\mathrm{samp}}$ for 4 of the blocks in the system. The true magnetisation of each block $ \langle \hat{M}_{B_i}(1) \rangle$ is estimated from the average of 10 Monte Carlo simulations. We also show the behaviour of the average deviation across all blocks $\delta \hat{M}$ (black line). (b) Plot showing $M(\underline{h})$ for different instances of $\underline{h}$ obtained using the survey-snapshot influence strategy (see main text Section \ref{three_block_structure}) as a function of $F_{\mathrm{samp}}$. Scatter points show average mangetisations obtained using 15 different instances of the survey-snapshot control. Black squares show the average value across these instances with the error bars indicating the standard error on the mean. Average mangetisations are computed using the average of 10 Monte Carlo chains with a burn-in time of $3 \times 10^5$ and tun time of $10^4$. All simulations performed for the case of $g=1$, $\beta = 8 \beta_c$ with the system initialised at (-1,-1,...,-1) on the Pokec social network.
	}
	\label{snapshot_performance}
\end{figure*}

Figure \ref{snapshot_mags} shows the behaviour of $\langle \delta \hat{M}_{B_x} \rangle$ as a function of $F_{\mathrm{samp}}$ for 4 of the blocks in the system. We also show the behaviour of $\delta \hat{M} = \frac{1}{q} \sum_{x=1}^q \langle \delta \hat{M}_{B_x} \rangle$, which represents the average of the absolute deviation across all the blocks. We observe that error shrinks relatively rapidly for values of $F_{\mathrm{samp}}$ between $10^{-3}$ and $10^{-2}$.

In order to explore the performance of the survey-snapshot control as we increase the sample size we draw a number of survey-snapshots and evaluate the survey-snapshot control for each of these. Figure \ref{snapshot_cons} illustrates how the consistency of the value of $M(\underline{h})$ obtained for different snapshots increases as a function of $F_{\mathrm{samp}}$. The average magnetisation across all instances of the survey-snapshot strategy remains stable with sample size. However, the variance in performance of the strategies decreases as $F_{\mathrm{samp}}$ increases indicating that we can achieve more consistent results by sampling more of the population.

For small values of $F_{\mathrm{samp}}$ we may sample single nodes for each block or a small number of nodes in the same state. In this case the estimated magnetisation for a block will be $\pm 1$. When this is the case the quantity used to capture the susceptibility $\beta(1-m_i^2)$ will be equal to zero. When $F_{\mathrm{samp}}$ is sufficiently small for this to occur for all blocks, all of the elements of the susceptibility vector will be zero meaning that we cannot meaningfully assign a control field using our approximation. Consequently, we do not consider values of $F_{\mathrm{samp}}$ significantly lower than $10^{-3}$. 

\clearpage

\subsection{Burn-in times for the Pokec network}

In this section we estimate the number of timesteps required for an Ising system to relax to a metastable solution on the Pokec social network. This relaxation time will depend on system parameters (e.g. $\beta$ and chosen external field $\underline{g}$) so we will focus on the choices used in the main text. 

\begin{figure*}
	\centering
	
	\subfloat[]{\label{Pokec_chains_1}\includegraphics[width=.4\textwidth]{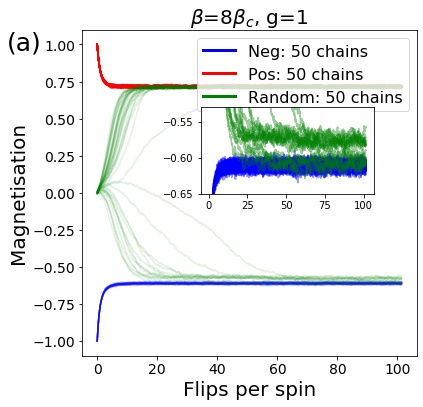}}
	\hspace*{4.0em}
	\subfloat[]{\label{Pokec_chains_2}\includegraphics[width=.4\textwidth]{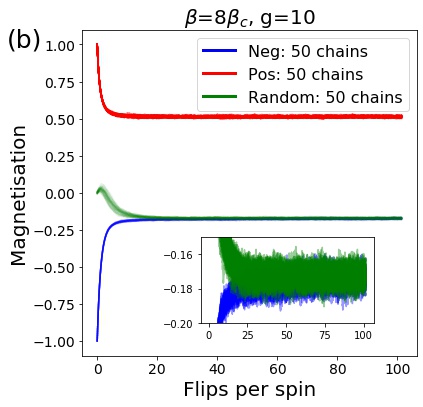}}
	\vspace*{-2.0em}
	\caption{\textbf{The burn-in time for the Pokec network with $\beta=8 \beta_c$ (relatively cold) is of the order of 10-50 flips per spin.} Plot showing average magnetisation as a function of the number of flips per spin ($\frac{T}{N}$) for the three different types of initial condition (positive, negative and random) described in text. Shown for $\beta=8\beta_{c}$ for a) $g=1$ and b) $g=10$. Each plot shows 50 Monte Carlo chains ran for $3 \times 10^6$ timesteps (approximately 100 flips per spin for the Pokec network) recording the average magnetisation every 1000 timesteps. Insets in a) and b) show a zoomed in view of the most negative metastable state.}
	\label{Pokec_robustness}
\end{figure*}

Let $\underline{s}_0$ represent the initial condition used in the MC simulations (method described in Section \ref{Ising_monte_carlo_sect}). We estimate the relaxation time for an Ising system on the Pokec social network by considering three different types of initial condition:
\begin{itemize}
	\item{Positive: $\underline{s}_0 = (1,1,...,1) = \mathbf{1}$  }
	\item{Negative: $\underline{s}_0 = (-1,-1,...,-1) = -\mathbf{1}$.}
	\item{Random: $\underline{s}_0 = (s_1,s_2,...,s_N)$, where $\mathbb{P}(s_i=1)=\mathbb{P}(s_i=-1)=\frac{1}{2}$. For this initial condition we use a different random condition for each Monte Carlo chain (whereas the others are fixed).}
\end{itemize}
The total external field acting on the system is given by: $\underline{g}=\underline{b}$ (i.e. $\underline{h}=0$), with $\underline{b}$ is set based on the age block of a node (see Section \ref{Pokec_external_sim}). We perform Monte-Carlo simulations on the Pokec network for each of the three initial conditions. Figure \ref{Pokec_robustness} shows the magnetisation as a function of the number of flips per spin for 50 different MC chains for each type of initial condition. For $g=10$ the system settles down to one of two different metastable solutions. For $g=1$ there are 3 metastable solutions in total. Two of these being relatively close together solutions with negative values of $M$.

This plot demonstrates that over the moderate timescale we do not see any transitions between different metastable solutions of the system. Consequently, that it is reasonable to consider moderate timescale control.

\section{Details of Code used for Numerical simulations}

The code used to perform simulations and generate the plots is available at: \href{https://github.com/MGarrod1/unobserved_spin_influence}{https://github.com/MGarrod1/unobserved\_spin\_influence}

This source code relies on two modules created by the author:
\begin{itemize}
	\item{\href{https://github.com/MGarrod1/ising_block_level_influence}{https://github.com/MGarrod1/ising\_block\_level\_influence}}
	\item{\href{https://github.com/MGarrod1/spatial_spin_monte_carlo}{https://github.com/MGarrod1/spatial\_spin\_monte\_carlo} }
\end{itemize}

The first is used to compute the magnetisation of Ising systems on networks under the mean-field approximation and the second is used to carry out the Monte Carlo simulations in order to validate the algorithms.

\end{document}